\documentclass[12pt]{article}
 \usepackage{amsmath}
 \usepackage{amssymb}
 \usepackage{bm}
 \usepackage{graphicx}
 \usepackage{graphics}
 \usepackage{epsfig}
 \usepackage{cite}

\title{Renormalization-group symmetries for solutions
of nonlinear boundary value problems}

\author{V. F. Kovalev$^{1}$, D. V. Shirkov$^{2}$ \\
\small\textit{$^{1)}$ Institute for Mathematical Modelling, Russian Academy of Sciences, } \\
\small\textit{ Miusskaya pl. 4a, 125047 Moscow, Russian Federation, e-mail: vfkv@orc.ru } \\
\small\textit{ $^{2)}$Joint Institute for Nuclear Research, ul. Joliot-Curie 6,} \\
\small\textit{  141980, Dubna, Moscow region, Russian Federation, } \\
\small\textit{ Physical Department, Lomonosov Moscow State University, } \\
\small\textit{ Vorob'evy gory, 119992 Moscow, Russian Federation }}

\date{}

\begin{document}
\maketitle

\begin{abstract}
Approximately 10 years ago, the method of renormalization-group symmetries entered the field of
boundary value problems of classical mathematical physics, stemming from the concepts of functional
self-similarity and of the Bogoliubov renormalization group treated as a Lie group of continuous
transformations. Overwhelmingly dominating practical quantum field theory calculations, the
renormalization-group method formed the basis for the discovery of the asymptotic freedom of strong
nuclear interactions and underlies the Grand Unification scenario. This paper describes the logical
framework of a new algorithm based on the modern theory of transformation groups and presents the
most interesting results of application of the method to differential and/or integral equation
problems and to problems that involve linear functionals of solutions. Examples from nonlinear
optics, kinetic theory, and plasma dynamics are given, where new analytical solutions obtained with
this algorithm have allowed describing the singularity structure for self-focusing of a laser beam
in a nonlinear medium, studying generation of harmonics in weakly inhomogeneous plasma, and
investigating the energy spectra of accelerated ions in expanding plasma bunches.

\end{abstract}
\newpage

\tableofcontents

\section{Introduction}

The paper presents materials illustrating the use and extensions of the concepts of functional
self-similarity and the Bogoliubov renormalization group in boundary value problems of mathematical
physics.

The (Lie transformation) group structure discovered by Stueckelberg and Petermann in the early
1950s in calculation results in renormalized quantum field theory and the exact symmetry of
solutions related to this structure were used in 1955 by Bogoliubov and one of the present authors
to develop a regular method for improving approximate solutions of quantum field problems, the
renormalization group (RG) method. This method is based on the use of the infinitesimal form of the
exact group property of a solution to improve a perturbative (that is, obtained by means of the
perturbation theory) representation of this solution. The improvement of the approximation
properties of a solution turns out to be most efficient in the presence of a singularity, because
the correct structure of the singularity is then recovered.

The most spectacular results obtained by the renormalization-group method in quantum field theory
were the discovery of the asymptotic freedom of non-Abelian gauge theories (Nobel Prize in 2004),
which led to the creation of quantum chromodynamics, and sketching the picture of the joint
evolution in energy of the three effective interaction functions (electromagnetic, weak, and
strong) in the Standard Model, which led to the speculative conjecture of a Grand Unification of
interactions and the possible instability of the proton. \par

Apart from this, the quantum field renormalization group provided a foundation (see K.G.~Wilson's
Nobel lecture, 1982) for the construction of an approximate semigroup in the investigation of phase
transitions in large spin lattices, the so-called Wilson renormalization group, which is widely
used in the analysis of critical phenomena.

In the present paper, we discuss the most interesting results obtained by the authors by extending
the RG concepts in quantum field theory to boundary value problems of classical mathematical
physics. The main achievement here was the development of a regular algorithm for finding
symmetries of the RG type by means of the modern theory of transformation groups. The existence of
such an algorithm eliminates the usual deficiency of the RG approach beyond the scope of quantum
field theory problems: finding the group property of solutions requires using special-purpose
methods of analysis, usually nonstandard, in each particular case. \par

It is notable that the algorithm of the construction of renormalization-group symmetries proposed
here can be applied to problems involving differential and integral equations, as well as linear
functional of the solutions. \par

We illustrate applications of the algorithm by examples from nonlinear optics, kinetic theory, and
plasma dynamics, including the problem of propagation and self-focusing of a wave beam in a
nonlinear medium (Sections \ref{rg-kcs} and \ref{rg-apr-ngo}), problems of the dynamics of a plasma
bunch and ion acceleration (Section \ref{plasma-dyn}), and the generation of harmonics in laser
plasma (Section \ref{plasma_harmonics}). There, the use of renormalization-group symmetries brought
about new exact and approximate analytic solutions of nonlinear physics problems, which allowed
describing the space structure of a self-focusing beam in a nonlinear medium in a realistic
setting, making significant progress in establishing relations between the intensity of the
harmonics generated by weakly inhomogeneous laser plasma in a strongly nonlinear regime and the
parameters of the radiation and the plasma, and finding, for the first time, the energy spectrums
of accelerated ions in the kinematic description of an adiabatic expansion of plasma bunches
consisting of several kinds of ions.

This paper is motivated by our desire to draw theorists' attention to a new and fairly general
algorithm based on applying the symmetry to an approximate solution for enhancing its approximation
power. The use of the group property (the symmetry) of a solution underlies both the
renormalization group method in quantum field theory and its analogue, the new renormalization
group symmetry algorithm in mathematical physics. \par

The universality of the renormalization-group ideas allows a unified approach to the analysis of
properties of solutions of various nonlinear problems and gives grounds for hopes that this method
can be efficiently used in other areas of contemporary physics.\par

As is known, this universality is a characteristic feature of another general method that
represents a solution as a 'path integral' (functional integral) and is widely used in quantum
mechanics, quantum field theory, the theory of large statistical systems, and turbulence theory.
\par

Classical mathematical physics deals with physical objects described by (ordinary or partial)
differential equations, which are nonlinear, or integrodifferential in most practically interesting
cases. Finding analytic solutions of such equations for arbitrary initial and/or boundary
conditions is impossible: normally, exact analytic solutions can only be found for initial and
boundary data of a special form; in other cases, we must content ourselves with approximate
solutions. The method of constructing a solution of a specific boundary value problem (BVP) is
usually peculiar to the equations of the particular problem under consideration.

In this paper, we present a method of investigation of analytic solutions based on the construction
and use of symmetries of a special form of BVP solutions, which we call symmetries of the
renormalization group kind or \textit{renormalization group} (RG) symmetries. We treat the notion
of 'symmetry' in the standard sense of continuous transformation groups: this means that a solution
of the BVP is transformed into another solution of the same BVP by a continuous transformation
group acting in the space of all the variables determining the solution. The attribute
'renormalization group' points to similarities existing between these symmetries and the symmetries
in quantum field theory related to the operation of renormalization of masses and charges (coupling
constants).

We note that a connection between symmetries and the problem of finding solutions of differential
equations was first established \cite{lie-anm-80,lie-col-22} by a Norwegian mathematician, Sophus
Lie (1842-1899), who showed that most results on the integration of ordinary differential equations
of various kinds can be obtained by a general method, subsequently called the group analysis of
differential equations. As one of the main ingredients of the theory of continuous groups, the
group analysis of differential equations allows classifying differential equations using the
language of symmetry groups, i.e., it produces a complete list of equations that can be integrated
(or such that their order can be reduced) by the group method and also suggests a \textit{regular}
procedure for finding these symmetries. Considerable progress in this area since the early 1950s
has led to new concepts and algorithms, and has also extended the range of possible applications of
the group analysis (see, e.g., monographs
\cite{ovs-bk-62,ovs-bk-78,ibr-bk-83,olv-bk-86,vin-bk-97,ibr-bk-99, fus-bk-90} and handbook
\cite{ibr-spr-94}), but it has not changed the general aims of the modern group analysis to develop
\textit{regular} methods of constructing and classifying solutions of nonlinear differential
equations on the basis of the symmetries of these equations. \par

In problems described by ordinary differential equations, the use of a symmetry group yields
general and particular solutions. In problems involving partial differential equations, which are
typical in mathematical physics, knowing a symmetry allows constructing particular solutions of a
BVP (invariant solutions, which are mapped into themselves by the group transformations, and
partially invariant solutions), with boundary data not known a priori and determined in the
construction of a specific solution. Because arbitrary boundary data are not normally invariant
under group transformations, the use of invariant solutions is generally considered inefficient for
the solution of BVPs.\par

Arguments underlying the renormalization group method in quantum field theory lead to a different
conclusion \cite{bsh-dan-55b}. This method uses the group property of a solution (expressed in
quantum field theory as a functional equation) for the enhancement of its approximation power.

Although the renormalization group method was originally formulated for quantum field problems, we
can explain its core idea by an example of a planar problem of radiation transfer
\cite{mna-dan-82,shr-umn-94}. We assume that the half-space $x > 0$ is filled with homogeneous
matter and a stationary stream of particles, characterized by a number $\alpha$, is falling on the
boundary $x=0$ of this medium. We consider the evolution of the number of particles in the stream
as it moves deeper into the medium. Let $\alpha_1$ be the number of particles in the stream at a
distance $x=l_1$ from the vacuum-matter interface and $\alpha_2$ the number of particles at the
distance $x=l_2=l_1+\lambda$ from the interface. Because the medium is homogeneous, the number of
particles moving inside at a distance $l$ from the interface is uniquely determined by some
function $A(l,\alpha)$ of the value of $\alpha$ at the interface and the distance $l$, i.e.,
$\alpha_1=A(l_1,\alpha)$ and $\alpha_2=A(l_2,\alpha)$. But the value $\alpha_2$ can also be
expressed as $\alpha_2=A(\lambda,\alpha_1)$ in terms of the same function $A(\lambda,\alpha_1)$ of
two variables, the distance $\lambda$ from the imaginary interface $x=l_1$ and the number of
particles $\alpha_1$ at this interface. Combining the two different definitions of $\alpha_2$, we
obtain the functional equation
\begin{equation}
 \label{funceq}
 A(l+\lambda,\alpha) = A(\lambda,A(l,\alpha))
\end{equation}
for $A(x,\alpha)$. The nature of the particles and the properties of the medium are irrelevant for
this consideration. Of course, solving the transport problem (i.e., an integrodifferential kinetic
equation), one can find the explicit (exact or approximate) form of the function $A(x,\alpha)$ in
each particular case, but the exact solution of the problem necessarily satisfies Eqn
(\ref{funceq}).
\par

A functional equation of form (\ref{funceq}) occurs naturally in considering a one-parameter group
$G$ of point transformations $T_a$ in the plane
\begin{equation}
 \begin{aligned}
 T_a: \qquad & \bar{x} = f(x,u,a)\,, \quad \bar{u} = g(x,u,a)\,,
 \quad f(x,u,0)= x\,, \quad g(x,u,0)= u \,,
 \end{aligned}
 \label{ta}
\end{equation}
mapping a point $P=(x,u)$ into another point ${\bar{P}}=(\bar{x},\bar{u})=T_{a}(P)$. We recall that
a set $G$ of invertible transformations $T_a$ forms a group if these transformations satisfy
several conditions: a) the set $G$ contains the identity transformation $T_0$; b) each $T_a$ has an
inverse transformation $T_{a^{-1}}$; c) the composition $T_b T_a$ of two transformations is also an
element of $G$:
\begin{equation}
 \label{defgroup}
 \begin{aligned} &
 T_{a^{-1}}(\bar{P})=P; \quad T_{0}(P)= P; \quad {\bar{\bar P}}
 \overset{def}{=} T_{b}(\bar{P})= T_b T_a(P)=T_{a+b}(P)\,. \\
 \end{aligned}
\end{equation}
The last condition in (\ref{defgroup}) can be expressed in terms of the functions $f$ and $g$ in
(\ref{ta}) as two functional equations:
\begin{equation}
 \label{funceq12}
 \begin{aligned} &
 f(x,u,a+b)=f(f(x,u,a),g(x,u,a),b), \\ &
 g(x,u,a+b)=g(f(x,u,a),g(x,u,a),b) \,. \\
 \end{aligned}
\end{equation}
It is known from the Lie theory that each continuous one-parameter group is fully determined by the
infinitesimal transformation
 \begin{equation}
  \begin{aligned} &
 \bar{x} = x + a \xi(x,u) + O(a^2) \,,
 & \bar{u} = u + a \eta(x,u) + O(a^2) \,, \\ &
 \xi(x,u)=(\partial f /\partial a)_{\big\vert a=0} \,,
 & \eta(x,u)=(\partial g / \partial a)_{\big\vert a=0}\,,
 \end{aligned}
\end{equation}
which is customarily expressed using the \textit{infinitesimal operator} (or generator)
 \begin{equation}
 \label{oper-def}
 X=\xi(x,u) \partial_{x} +
   \eta(x,u) \partial_u
 \end{equation}
of the group. Finite transformations of a continuous group are uniquely determined by the
infinitesimal generator by means of the Lie equations, which are the characteristic equations for
the first-order partial differential equation associated with (\ref{oper-def}),
 \begin{equation}
 \label{lieeq-pt}
 \begin{aligned}
 \frac{d {{\overline{x}}}}{d a}  & =
 \xi(\overline{x},\overline{u})\,, \quad
 {\overline{x}}{\vert}_{a=0} =x\,, \quad
 \frac{d {{\overline{u}}}}{d a}  & = \eta
  (\overline{x},\overline{u})\,, \quad
 {{\overline{u}}}{\vert}_{a=0} =u\,. \\
 \end{aligned}
 \end{equation}
For the radiation transfer problem under consideration, we have $f=x$, $g=A(\alpha,\lambda)$,
$a=\lambda$, and $u=\alpha$, and functional equation (\ref{funceq}) coincides with the second
equation in (\ref{funceq12}) (the first equation there is an expression of the obvious additive law
of the transformation of the coordinate $\bar{x}=\lambda$), and the group generator is given by
 \begin{equation}
 \label{oper-trans}
 X = \partial_{x} +
   \eta(\alpha) \partial_\alpha\,, \quad
   \eta(\alpha)\equiv
   \partial_\lambda A(\alpha,\lambda){\big\vert}_{\lambda=0} \, .
 \end{equation}
In accordance with (\ref{lieeq-pt}), to find $A$ at a large distance from the boundary, i.e., for
large values of the parameter $\lambda$, we must know the behavior of
${\bar{\alpha}}=A(\alpha,\lambda)$ in a thin boundary layer, as $\lambda \to 0$, i.e., we must in
fact know the derivative of this function at the boundary. This information can usually be
extracted from an approximate solution provided by the perturbation theory. Next, integration of
the Lie equations yields formulas for finite transformations:
 \begin{equation}
 \label{solution-lieeq}
 \begin{aligned}
 \Psi({\bar{\alpha}})  = \Psi(\alpha)  +  \lambda \, ,
 \quad \Psi(\alpha)  = \int\limits^{\alpha} {\rm d} a / \eta(a) \,, \quad
 {\bar{x}} = \lambda \,.
 \end{aligned}
 \end{equation}
Assuming that $\Psi$ has the inverse function $\Psi^{-1}$, we find solutions of functional equation
(\ref{funceq}) in the general form:
 \begin{equation}
 \label{solution-funceq}
 \begin{aligned}
 {\bar{\alpha}} \equiv A(\alpha,\lambda)  =\Psi^{-1}( \Psi(\alpha)  +  \lambda) \, ,
 \quad  {\bar{x}} = \lambda \,.
 \end{aligned}
 \end{equation}
These constructions are the essence of the renormalization group method. We now present two
examples of implementing this method. \par

We consider a medium absorbing particles in proportion to their number; from the perturbation
theory, we know the approximate solution
 \begin{equation}
 \label{transf-pert1}
 A_{pt}(\alpha,\lambda) \approx \alpha - \nu \alpha \lambda \,, \quad \nu=const.
 \end{equation}
Calculating the coordinate $\eta(\alpha)= - \nu \alpha$ of group generator (\ref{oper-trans}) with
the help of (\ref{transf-pert1}) and using it in relations (\ref{solution-lieeq}), we obtain
solution (\ref{solution-funceq}),
 \begin{equation}
 \label{transf-rg1}
 A_{rg}(\alpha,\lambda) = \alpha \exp \left( - \nu \lambda \right) \,,
 \end{equation}
which is valid in the entire space filled with matter, up to $x \to \infty$.
\par

We now assume that the absorption has a nonlinear mechanism, with the absorption coefficient
proportional to the stream of particles, $\nu(\alpha)=\beta \alpha$, where $\beta=$ const. In a
thin boundary layer, we then have
 \begin{equation}
 \label{transf-pert2}
 A_{pt}(\alpha,\lambda) \approx \alpha - \beta \alpha^2 \lambda \,,
 \end{equation}
and the use of (\ref{transf-pert2}) in (\ref{solution-lieeq}) yields a solution of
(\ref{solution-funceq}) in the form of the sum of a geometric progression:
 \begin{equation}
 \label{transf-rg2}
 A_{rg}(\alpha,\lambda) = \frac{\alpha}{1+ \beta \alpha \lambda} \,.
 \end{equation}
This result, similarly to the previous one, holds in the entire subspace $x>0$ and, in particular,
describes the asymptotic behavior of the permeating stream as $x \to \infty$.
\par

The efficiency of the (renormalization) group approach in the above examples shows itself in the
following fact: using information about the behavior of a solution in a neighborhood of the
vacuum-medium interface, we obtain explicit expressions   for   the   solution over   the   entire
interval $0 \leqslant  x \leqslant \infty$. We note that if we expand expression (\ref{transf-rg2})
in a power series in the particle number density, % $\alpha$
i.e., return to the perturbation theory, then in each order in $n$, we obtain expressions
increasing in proportion to $\lambda^n$, which is a distorted representation for the asymptotic
form of the solution. The advantage of the renormalization group method is the recovery of the
actual structure of the solution, consistent with functional equation (\ref{funceq}), which is
distorted by perturbation theory approximations.{\footnote{ We note that formulas
(\ref{transf-rg1}) and (\ref{transf-rg2}) can also be obtained in transfer theory by other methods,
for instance, by solving the kinetic equation; however, the method described here, based on the use
of group differential equations, is the simplest. Moreover, in several important cases, this is the
only possible method: results obtained with this method are unattainable in other ways.}} \par

In the case of the planar problem of radiation transfer, the transparency of the renormalization
group method is a consequence of taking account of the symmetry properties of solutions (i.e., of
the functional equation for them) in the actual configuration space. The RG transformation of the
particle number density in moving deeper into the medium is related to a shift in the spatial
coordinate. \par

Returning to the renormalization group method in its original (quantum field) formulation
\cite{bsh-dan-55b,bsh-ncm-56,bsh-jetf-56,bsh-bk-80}, which is also called the Bogoliubov
renormalization group,\footnote{This term was introduced in \cite{shr-umn-94} to distinguish the
Bogoliubov RG from different constructions also called renormalization groups in some other areas
of physics. They are briefly listed, e.g., in \cite[Sec.3]{shr-umn-94}.} we note that it is based
on a functional equation that in the simplest case has the same form as (\ref{funceq}) after the
substitution $x \to \ln t$, such that the RG shift transformation of a spatial variable in transfer
theory corresponds to the transferred momenta rescaling in quantum field theory; the quantity
$\bar{\alpha}$ is called the invariant (effective) coupling function in this theory. In particular,
a solution of form (\ref{transf-rg2}) with $\lambda=\ln x$ occurs in quantum field calculations in
the one-loop approximation. If a more advanced perturbation-theory approximation is used, which
differs from (\ref{transf-pert2}) by the presence of terms cubic in $\alpha$, which corresponds to
the two-loop approximation in quantum field theory, then the RG-improved solution can be found from
an equation similar to (\ref{solution-funceq}) that is unsolvable in elementary functions
\cite{sol-tmf-99}. It is usually solved iteratively, using the one-loop approximation of the RG
expression.\par

The comparison of the RG-improved solution found in the two-loop approximation with the result
obtained in the one-loop approximation reveals a characteristic feature of the renormalization
group method: we can progressively improve the accuracy, which is an indication of the stability of
the asymptotic behavior of the solution. Similarly, in the perturbation theory, we can also take
higher-order corrections into account, which successively improves the corresponding RG
solutions.\par

Thus, the procedure for the systematic (successive) improvement of the system of approximate
solutions found in quantum field theory in the perturbation theory with respect to a known small
parameter is quite similar to the above. This improvement of the approximation properties is most
significant in the neighborhood of a singularity of the solution. In the quantum field context,
these are singularities in the infrared (see \cite{bsh-ncm-56,bsh-jetf-56,log-jetp-56}) and
ultraviolet domains. The latter include the most spectacular result obtained with the help of the
RG method, the discovery of the asymptotic freedom of non-Abelian gauge theories \cite{af-73}. \par

The above examples of the use of the renormalization group method for improving and refining the
approximation properties of perturbative solutions are based on a functional equation of the
simplest form, with one or two independent and one dependent variable. But the number of
independent and dependent variables in the problem is often larger than this minimal set.\par

For example, a version of functional equation (\ref{funceq}) with $x=\ln t$ corresponds to a
massless model with one coupling constant in quantum field theory. We can make this model more
involved in two ways. First, the number of arguments defining the effective coupling
${\bar{\alpha}}$ can be increased. For instance, the field model under consideration can contain
one or several masses (e.g., as in quantum chromodynamics); in that case, the effective coupling
acquires a dependence on several mass variables with the corresponding transformation laws, with
the result that the group transformations and the functional equation change their form. Second,
the number of functional equations can be larger, which corresponds to a quantum field model with
several coupling constants. This means that we now consider a group of continuous transformations
of independent variables $x=\{x^1, \ldots , x^n \}$ and dependent variables $u=\{u^1, \ldots , u^m
\}$ with infinitesimal operator (\ref{oper-def}) in the space $R^{n+m}$, and the coordinates of
generator (\ref{oper-def}) are vectors $\xi=\{\xi^1, \ldots , \xi^n \}$ and $\eta=\{\eta^1, \ldots
, \eta^m \}$; the corresponding contributions to the infinitesimal operator must be understood as
the result
of the contributions of the individual variables. %, e.g., $\xi
 % \partial_x \equiv \xi^1 \partial_{x^1} + \ldots + \xi^n \partial_{x^n}$.
With an increase in the number of arguments of the function to be governed by the functional
equation and an increase in the number of the equations themselves, finding the group property of
the solution that can be expressed by a functional equation (if we use the original formulation of
the renormalization-group method~\cite{bsh-ncm-56}) requires a special and often nontrivial
analysis in each particular case (see, e.g., the discussion in \cite{shr-umn-94,gol-pre-96}); from
the algorithmic standpoint, this is a deficiency of the RG techniques outside the quantum field
theory.
\par

To overcome this deficiency in extending the RG concepts to problems of mathematical physics,
another special RG algorithm was developed (see \cite{kov-tmf-89,shkkp-rg-91} and also reviews
\cite{kov-jmp-98};\cite[p.232]{ksh-phr-01};\cite{ksh-proc-04,ksh-rg-05}). It has the same aim of
finding an improved solution (in comparison with the initial approximate solution) as the algorithm
of Bogoliubov's RG method, but in finding symmetries of a solution of a BVP it uses a scheme of
calculations similar to that of the modern group analysis. This feature explains the term 'RG
symmetry' \cite{kov-jmp-98}. \par

In this paper, we describe the RG algorithm in mathematical physics and illustrate its capabilities
by various examples of BVPs. The paper is organized as follows. In Section~\ref{illustration}, we
explain the core ideas of the RG algorithm using the example of the construction of an RG symmetry
for a solution of a BVP for the Hopf equation. Sections~\ref{rgs-loc} and \ref{rgs-nlc} illustrate
different approaches to the construction of RG symmetries; furthermore, in Section~\ref{rgs-loc},
we consider several progressively more complicated problems obtained by modifying and supplementing
the Hopf equations in Section~\ref{illustration}. Section~\ref{rgs-nlc} follows the same logic, but
we supplement the presentation there with a discussion of nonlocal problems, which are not
necessarily connected with the ones in Section~\ref{rgs-loc}. The scope for possible applications
of the RG algorithm and a brief list of results obtained with its use are presented in
Section~\ref{conclusion}.

%\end{document}

\section{The renormalization-group algorithm in mathematical physics \label{illustration}}

We preface the description of the RG algorithm with the following simple argument. It is known that
if we treat all the variables (independent or dependent in the standard sense) involved in a
differential equation and their derivatives (called differential variables in group analysis) as
independent, then the differential equation can be regarded as an algebraic relation for these
variables. In the case of one equation, this relation describes a 'surface' in the extended space
of all the variables involved in the equation (if there are several equations, then we speak of a
manifold), and each solution of the equation defines a 'line' on this surface. The projection onto
the $\{x,u\}$ 'plane' defines a family of curves, one of which passes through the 'point'
$\{x_0,u_0\}$ corresponding to the boundary condition of the BVP in question. \par

Transformations of the group $G$ move points on the surface (the manifold) along this surface, and
therefore the equation preserves its form in the transformed variables and each solution of the
equation is taken into another solution. A transformation $T_a$ from the group $G$ maps a point on
the plane $\{x,u\} \in \mathbb{R}^{n+m}$ into a point $\{\bar{x},\bar{u}\}$, and the geometrical
locus of these points is a continuous curve (a trajectory of the group $G$) passing through
$\{x,u\}$. The locus of images $T_a(\{x,u\})$ is also called the $G$-orbit of the point $\{x,u\}$.
In the general case, the motion along a group trajectory corresponds to the transition from one
curve in the family to another, that is, to a 'multiplication' of solutions.\par

Returning to the renormalization-group point of view, we consider only the group transformations
under which points on the curve passing through $\{x_0,u_0\}$ are moved along this curve. This
means that the solution of the BVP is the RG orbit of the point $\{x_0,u_0\}$ (of the boundary
manifold in the general case) and is an invariant RG manifold (similarly to the invariant charge in
quantum field theory \cite{bsh-bk-80}). We use the infinitesimal version of this property in our
construction of the RG symmetry.\par

The group property of a solution of a BVP manifests itself as follows: instead of the boundary
point $\{x_0,u_0\}$ parameterizing the solution, we can take another point in this curve related to
it by an RG transformation. This 'universality' of the solution of a BVP under a change of the way
of parameterization is called 'functional self-similarity' \cite{shr-dan-82}. To find RG
transformations that map a solution of a BVP into a solution of the same BVP, we use the fact that
a physical problem is formulated in terms of differential (integrodifferential) equations whose
symmetries can be found by the techniques of group analysis. \par

We now illustrate the characteristic features of the algorithm for constructing an RG symmetry by
an example of a BVP for the Hopf equation \cite{ksh-rg-05}, which is widely used in physics for the
description of the initial perturbations at the nonlinear stage of their evolution:
\begin{equation}\label{hopf}
 \partial_t v + v \partial_x v = 0\,, \quad v(0,x)= \epsilon U(x)\,,
 \end{equation}
where $U$ is an invertible function of $x$ and the parameter $\epsilon$ defines the 'amplitude' of
the initial perturbation 'at the boundary' $t = 0$. For a very small distance $t \ll 1/\epsilon $
from the boundary, the solution of problem (\ref{hopf}) given by the perturbation theory is a
segment of a power series,
 \begin{equation}
 v = \epsilon  U - \epsilon^2 t U \partial_x U + O\left( t^2 \right) \,,
 \label{hopf-pt} \end{equation}
but this form becomes inapplicable for large $t$. The RG symmetry allows improving the perturbative
result and recovering the correct behavior of the solution in a neighborhood of a singularity (when
such a singularity occurs for some values of $t$). \par

In constructing an RG symmetry, the algorithm uses the symmetry group of the BVP equations. The
boundary data defining a particular solution are involved in RG transformations by extending the
space of the variables on which the group acts. In the case of BVP (\ref{hopf}), this space
involves three independent variables, $x=\{t,x,\epsilon \}$. It is convenient to write differential
equation (\ref{hopf}) for the function $u=v/\epsilon$ introduced such that the 'amplitude'
$\epsilon$ is carried over from the boundary condition to the differential equation:
\begin{equation}\label{hopf-eq}
 \partial_t u + \epsilon u \partial_x u = 0\,, \quad u(0,x)= U(x)\,.
\end{equation}
The general element of the transformation group $G$ for Eqn (\ref{hopf-eq}) (for the basic manifold
in the general case) can be found by means of the standard Lie techniques (see, e.g.,
\cite{ovs-bk-78}); it is given by a combination of four infinitesimal operators,
\begin{equation} \label{gen-hopf}
  \begin{aligned} &
 X=\sum\limits_{i} X_i \, , \quad
  X_1=\psi^{1} \left( \partial_{t} +
 \epsilon u \partial_{x} \right) \,, \quad
 X_2=\psi^{2} \partial_{x} \,, \\ &
 X_3=\psi^{3} \left(x \partial_{x} + u \partial_{u} \right) \,, \quad
 X_4=\psi^{4} \left(\epsilon \partial_{\epsilon}
 + x \partial_{x} \right) \,,
 \end{aligned}
 \end{equation}
where $\psi^i$ ($i=2,3,4$) are arbitrary functions of $\epsilon$, $u$ and $x-\epsilon u t$ and
$\psi^1$ is an arbitrary function of all the group variables $\{t,x,\epsilon,u\}$. We now use the
RG invariance condition for a particular solution of BVP (\ref{hopf-eq}) defined by the relation
\begin{equation}\label{hopf-inv}
 S \equiv u-W(t,x,\epsilon) = 0
\end{equation}
with the function $W$ that is unknown at this point; in other words, we check that the
RG transformation maps the solution of the BVP into the same solution. In the
infinitesimal form, this condition can be written as
 \begin{equation} \label{invarcond-hopf}
 {X S}_{\big\vert {[S]}} \equiv
 \psi^3 (W-x\partial_x W) - \psi^2 \partial_x W
 - \psi^4 (\epsilon \partial_\epsilon W +x \partial_x W) = 0 \,,
 \end{equation}
where $\vert_{[S]}$ means that the result of the action of the operator is taken on the manifold
defined by the equation $S=0$ and all its differential consequences. The term containing $\psi^1$
is absent in (\ref{invarcond-hopf}) because it is proportional to $\partial_t W + \epsilon W
\partial_x W$, which vanishes identically on solutions of Eqn (\ref{hopf-eq}). Condition (\ref{invarcond-hopf}) holds
for all $t$, and for $t \to 0$ in particular, when $W$ is replaced by the approximate solution
\begin{equation}
 W = U - \epsilon t U \partial_x U + O \left( t^2 \right)
 \label{hopf-pt1}
 \end{equation}
obtained in the framework of perturbation theory (\ref{hopf-pt}). In this limit, Eqn
(\ref{invarcond-hopf}) yields a relation for the functions $\psi^i$ $(i=2,3,4)$, which extends in
the obvious fashion to $t \neq 0$:
\begin{equation} \label{restrict-hopf}
 \psi^2 = - \chi ( \psi^3 + \psi^4) +(u/\partial_{\chi} U) \psi^3 \,,
 \quad \chi = x - \epsilon u t \,, \end{equation}
where the derivative $\partial_{\chi} U$ must be expressed in terms of $\chi$ or $u$  in accordance
with the boundary conditions. Using (\ref{restrict-hopf}) in (\ref{gen-hopf}), we arrive at a group
of a smaller dimension with the infinitesimal operators
\begin{equation} \label{rg-hopf}
 \begin{aligned} &
 R=\sum\limits_{i} R_i \, , \quad
 R_1=\psi^{1} \left( \partial_{t} + \epsilon u \partial_{x} \right) \,, \\ &
 R_2= u \psi^{3} \left[ \left( \epsilon t +1/\partial_{\chi}U \right) \partial_{x}
 + \partial_{u} \right] \,, \quad
 R_3= \epsilon \psi^{4} \left( t u \partial_{x}
 + \partial_{\epsilon} \right) \,.
 \end{aligned} \end{equation}
The above procedure reducing (\ref{gen-hopf}) to (\ref{rg-hopf}) is \textit{the restriction of
group} (\ref{gen-hopf}) \textit{on a particular solution}, and the set of operators $R_i$ in
(\ref{rg-hopf}) describes the required RG symmetry. We obtain the solution of the BVP with the use
of the corresponding Lie equations (similar to (\ref{lieeq-pt})) for any generator in
(\ref{rg-hopf}). Without loss of generality, we can take the generator $R_3$ with $\epsilon
\psi^{4}=1$ to obtain the finite RG transformations
 \begin{equation}\label{trans-hopf}
 x^{\prime} = x + a t u \,, \quad
 \epsilon^{\prime} = \epsilon + a \,, \quad
 t^{\prime} = t \,, \quad u^{\prime} = u \,, \end{equation}
where $a$ is the group parameter, $t$ and $u$ are invariants, and the transformations of $\epsilon$
and $x$ are translations, which in addition depend on $t$ and $u$ for the $x$ variable. For
$\epsilon = 0$, in view of (\ref{hopf-eq}), the variables $x$ and $u$ are related by $x = H(u)$,
where $H(u)$ is the function inverse to $U(x)$. Eliminating $a, t$, and $u$ from (\ref{trans-hopf})
and dropping the dashes in our notation for the variables, we obtain the required solution of BVP
(\ref{hopf-eq}) in implicit form [similar to the implicit form of the solution of functional
equation (\ref{solution-funceq})]:
 \begin{equation}
 x - \epsilon t u = H(u) \,.
 \label{solution-hopf}
 \end{equation}
In effect, this is the improved perturbation theory solution (\ref{hopf-pt}), which can be used not
only for small $ t \ll 1/\epsilon$ (of course, under the condition that (\ref{solution-hopf})
defines $u$ uniquely). Depending on $H(u)$, this solution either indicates the correct asymptotic
behavior as $t\to \infty$ or gives the correct description of the solution in the neighborhood of
finite values $t\to t_{sing}$. One example of the first option is the solution of the BVP for the
linear function $U(x) = x$. This yields the expression $v=\epsilon x (1+\epsilon t)^{-1}$, which
remains finite as $t \to \infty $, similarly to the solution of (\ref{transf-rg2}). For the second
option, we can select, for instance, a sine wave $U(x)= - \sin x$ at the boundary. Then solution
(\ref{solution-hopf}) describes the well-known distortion of the initial profile of a sine wave,
transforming it into a saw-tooth shape \cite[Ch.6, \S1]{rud-bk-75}, with a singularity forming at a
finite distance $t_{sing}=1/\epsilon$ from the boundary. We note that for finding solution
(\ref{solution-hopf}) of the BVP, we use \textit{only} the known symmetry of the solution and the
corresponding perturbation theory (PT). \par

The above example of the construction of RG symmetries illustrates the general algorithm, whose
detailed description in relation to BVPs for differential equations can be found, e.g., in reviews
\cite{kov-jmp-98,ksh-phr-01}, and whose generalization to nonlocal problems is presented in
\cite{ksh-rg-05,ksh-proc-04}. We can schematically express the implementation of the RG algorithm
as a sequence of four steps (see the figure):

\hspace{-0.5cm}
\begin{minipage}[b]{.40\linewidth}
 \begin{itemize}
 \item[\textbf{(I)}] constructing the \\ basic \textit{manifold} $\cal{RM}$;
 \item[\textbf{(II)}] finding a \textit{symmetry group}  $G$ admitted by $\cal{RM}$;
 \item[\textbf{(III)}] \textit{restricting the symmetry group} $G$ on a particular solution of the BVP  and finding the RG
                       symmetry (RGS);
 \item[\textbf{(IV)}] finding an \textit{analytic solution} corresponding to the RG symmetry.
\end{itemize}
 \end{minipage} %\hspace{0.65cm} \hfill
\begin{minipage}[b]{0.59\linewidth}
 \centering\epsfig{figure=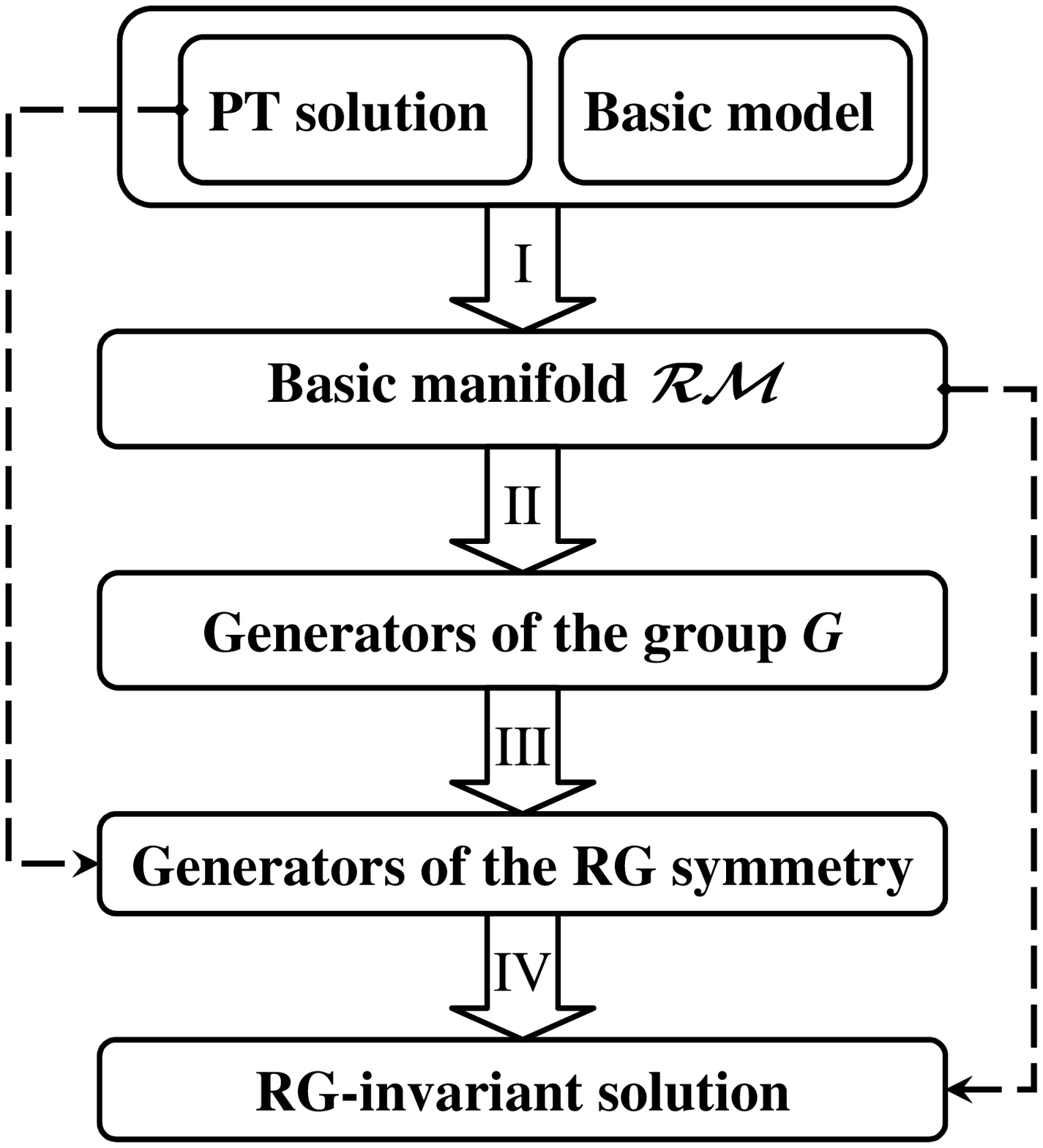,width=\linewidth } \\
 \small{\textbf{Figure.} The scheme of the RG algorithm}
 \end{minipage}\medskip

\vspace*{0.3cm}

A characteristic feature of the procedure of constructing the RG symmetry is the multivariance of
step \textbf{(I)}, whose aim is to have the parameters participating in the equations and the
boundary conditions of the problem and determining the solution somehow involved in
transformations. The choice of a concrete realization of the first step is most usually governed by
the form of the basic equations and the corresponding boundary conditions on the one hand and by
the form of the approximate PT solution on the other. This multivariance, which is a feature of
step \textbf{(I)} alone, is aimed at covering a possibly broader spectrum of problems to be
investigated by the method. The subsequent steps are carried out in the framework of well-developed
group methods. \par

This multivariance is also seen in the above simple example of a BVP for the Hopf equation.
Underlying our construction of the RG symmetry for BVP (\ref{hopf-eq}) was the most obvious option:
constructing the RG symmetry from the point symmetry group of the Hopf equation in the space
extended by incorporating the parameter $\epsilon$ into the set of independent variables. This way
of constructing the basic manifold $\cal{RM}$ is not the only possible one. \par

We could also construct the RG symmetry for BVP (\ref{hopf-eq}) using an additional differential
constraint compatible with the boundary conditions and the basic equations.\footnote{Here, we do
not detail the construction of such a differential constraint. As an example, we note the use of
the invariance condition for the basic equation under so-called higher (or Lie-B\"acklund)
symmetries rather than point symmetries. In contrast to the coordinates of an infinitesimal
generator of a point symmetry group, the coordinates of a generator of a higher-symmetry group in
addition to independent and dependent variables, also depend on higher derivatives. Expressing the
invariance condition under a group of higher symmetries in the infinitesimal form, we obtain the
required differential constraint (see \cite{kov-jmp-98} for the details).} For instance, if the
initial conditions in (\ref{hopf-eq}) are given by the linear function $U(x) = x$, then we can take
the differential constraint $\partial_{xx}u=0$. Next, we calculate the RG symmetry of BVP
(\ref{hopf-eq}) taking the basic manifold $\cal{RM}$ to be the system obtained by combining this
constraint and the Hopf equation. The admissible group $G$ for the manifold $\cal{RM}$ in so doing is
different from (\ref{gen-hopf}), but the form of solution (\ref{solution-hopf}) is the same. Other
examples of the implementation of step \textbf{(I)} of the algorithm can be found in
\cite{kov-jmp-98}.

\section{Renormalization-group symmetries in local problems of mathematical physics \label{rgs-loc}}

The example of the construction of the RG symmetry for Hopf equation (\ref{hopf-eq}) demonstrates
that a particular form of the realization of the general scheme of the RG algorithm depends on the
form of the equations in the BVP, as well as on the way the boundary data are specified. Since the
construction of the RG symmetry proceeds by restricting the symmetry group $G$ of the basic
manifold [step \textbf{(III)}], the RG usually has a smaller dimension than $G$. For instance, in
the case of BVP (\ref{hopf-eq}), the symmetry group $G$ is defined by the four generators $X_i$,
and the RG is defined by the three generators $R_i$. It is obvious that for the construction of the
RG symmetry, it is desirable to have a maximal group $G$. However, the more complicated the basic
equations are, the narrower the admissible transformation group typically is. For instance, if the
term $\nu u_{xx}$ accounting for dissipation is added to the Hopf equation, then after the change
of variables $u_x=w$, we obtain the modified Burgers equation. For this equation, the admissible
symmetry group is infinite-dimensional, but it is now characterized by a single arbitrary function
instead of four functions for the Hopf equation, and after the reduction procedure, we obtain a
finite-dimensional (8-dimensional) RG \cite{kov-lga-94}. \par

It is also possible that the RG symmetry cannot be constructed using a point symmetry group for the
basic manifold alone because restricting on a particular solution yields a zero-dimensional group.
In this case, we must either modify (and simplify) the system of equations used for the description
of the physical process or use other symmetries in addition to Lie symmetries for constructing the
RG. \par

We now demonstrate various approaches to the construction of the RG symmetry for the BVP obtained
by complicating the problem in (\ref{hopf-eq}), which was our example of the construction of the RG
symmetry in Section~\ref{illustration}.

\subsection{Renormalization-group symmetry in nonlinear plasma theory \label{plasma_harmonics}}

We consider the following problem, which was historically the first example of a successful
application of the RG algorithm. This is the interaction of $p$-polarized electromagnetic radiation
with a frequency $\omega$ and a 'moderate' (by today's standards) intensity, with inhomogeneous
plasma \cite{kov-tmf-89}. This interaction is described by a system of 2-dimensional nonstationary
differential equations (the equations of the collisionless hydrodynamics of electron plasma with a
self-consistent electromagnetic field) for six functions: the components $B_z$ and $E_x$, $E_y$ of
the magnetic and the electric fields, two components $V_x$, $V_y$ of the velocity of the electrons,
and their density $n$; these functions depend on three variables: the coordinates $x$ and $y$ and
time $t$. Our aim is to obtain an approximate analytic solution of this system of equations in an
arbitrary order of nonlinearity, without confining ourselves to the perturbation-theory framework.
\par

For an arbitrary ion density function $n^i(x)$, the basic system of equations admits only a
finite-dimensional point transformation group, the group of translations along the $t$ and $y$
axes. If the ion density is a constant, $n^i \equiv N =$ const, then we also have the group of
translations along the $x$ axis and the group of simultaneous rotations in the three planes defined
by the coordinates $\{x,y\}$ and the corresponding $x$ and $y$ components of the velocity of the
electrons and of the electric field. Thus, regarding the original equations as the manifold
$\cal{RM}$, we obtain a fairly narrow admissible group, which does not allow finding the required
RG symmetry. \par

To construct a manifold $\cal{RM}$ allowing a wider point transformation group, we use the fact
that the leading contribution to nonlinear effects of the interaction of the electromagnetic wave
with the inhomogeneous (in $x$) plasma under consideration here comes from the components of the
electric field and the velocity of the electrons that are directed along the density gradient.
Furthermore, due to the natural smallness parameters (the smooth inhomogeneity of the ion density
 % $n^i(x)$ %
along the $x$ axis and the small angle of incidence $\theta$ of the laser beam on the plasma), the
dependence of these components on the $y$ coordinate, which is transverse to the density gradient,
is smoother than their dependence on $x$ in the neighborhood of the plasma resonance. Hence, in the
construction of $\cal{RM}$, in the full system of 6 original equations, we can single out a simpler
system of two one-dimensional nonlinear partial differential equations for the $x$ components $E_x$
of the electric field and $V_x$ of the velocity of the electrons in the neighborhood of the plasma
resonance:
 \begin{equation}\label{twoeq}
 \begin{aligned}
 &
 \omega \partial_{\tau} v
    + a v \partial_x v -p  = 0  \,, \quad
 \omega \partial_{\tau} p
    + a v \partial_x p +\omega_L^2 v = 0  \,, \quad
    & \tau \equiv \omega t - (\omega y /c ) \sin \theta \,. \\
 \end{aligned}
 \end{equation}
Here, $v$ and $p$ are the respective quantities $V_x$ and $E_x$ normalized by the parameter $a$,
the parameter $a\propto \sqrt{q}$ is determined by the radiation flux $q$ on the plasma and the
linear transformation coefficient, $\omega_L (x)$ is the plasma frequency (for the fixed ion
density), and $c$ is the speed of light. \par

The infinite-dimensional point transformation group in the space of 5 variables $\{\tau, x, a, $
$v, p\}$ admitted by (\ref{twoeq}) is defined by an infinitesimal operator, which is a sum of three
operators:
 \begin{equation} \label{group-el}
 \begin{aligned}
 & X =\sum_{i=1}^{3}X_i\,, \quad X_1  =\mu_1 Y\,, \quad X_2=\mu_2 \partial_x
     +\frac{1}{a}Y(\mu_2)\partial_v +\frac{1}{a} Y^2(\mu_2)\partial_p \,, \\
 & X_3 = \frac{\mu_3}{a}(a\partial_a-v\partial_v-p\partial_p)\,, \quad
 Y  =\omega\partial_{\tau}+av\partial_x+p\partial_v
     -\omega_L^2 v\partial_p \,. \\
 \end{aligned}
 \end{equation}
Each of these three operators involves an arbitrary function $\mu_i$ of the group variables, where
$\mu_2$ and $\mu_3$ satisfy the differential constraints
 \begin{equation} \label{restrict-2}
 Y^3(\mu_2)+Y(\omega_L^2\mu_2)=0, \qquad Y(\mu_3)=0.
 \end{equation}
To specialize the functions $\mu_1$, $\mu_2$, and $\mu_3$  entering the coordinates of the operator
$X$, we use the procedure of restriction of the point transformation group~(\ref{group-el}) on an
approximate (in powers of $a$) solution of the BVP. We can construct this solution such that the
zeroth approximation to the functions $v$ and $p$ is found by solving the linearized system of the
original six equations endowed with the corresponding boundary conditions (an electromagnetic wave
falling on plasma from the vacuum) and by the selected density profile $n^i(x)$ in the plasma
resonance region; corrections to this solution that are proportional to $a$ arise after the
linearization of system~(\ref{twoeq}). The verification of the RG-invariance conditions [similar to
(\ref{invarcond-hopf})] for this approximate particular solution determines our choice of the
functions $\mu_1=0$, \ $\mu_2=-p/\omega^2$, and $\mu_3=1$ and yields the required RG-symmetry
operator (where the first relation in (\ref{restrict-2}) holds with the substitution $\omega_L^2$
$\to$ $\omega^2$):
 \begin{equation}
 \label{rg-el-gen}
 R = X_2 + X_3 = - (p / \omega^2) \partial_x + \partial_a.
  \end{equation}
The quantities $\tau$, $v$ and $p$ are invariants of the RG transformations with infinitesimal
operator (\ref{rg-el-gen}), and the transformation of the $x$ variable defined by the solution of
the Lie equation for (\ref{rg-el-gen}) exhibits a linear dependence on the parameter $a$:
\begin{equation}
 \label{rg-x-trans}
 x = \eta - (p /\omega^{2}) a.
  \end{equation}
The group composition law for $x$ can be easily deduced from the functional equation of form
(\ref{funceq}) with the substitutions $A \to x$, $l\to a_2$, $\lambda \to a_1$, and $\alpha \to
\eta$. We note that in contrast to the transfer theory problem, the group parameter here is not an
independent variable involved into the equation but the parameter $a$ imported into the equation
from the boundary conditions. \par

The solution of Eqn (\ref{twoeq}) constructed with the help of (\ref{rg-el-gen}) is given by
 \begin{equation}\label{solution-el}
 \begin{aligned} &
  a p /\omega^2 \Delta  = - \varepsilon  \left(
  f_1 \sin \tau + f_2 \cos \tau \right) \,, \quad
  \varepsilon  \equiv (q/q_0)^{1/2}  \ ,   \\ &
  a v /\omega \Delta =  \varepsilon \left(
  f_1 \cos \tau - f_2 \sin \tau \right)\,, \quad
  x = \eta + \varepsilon \left( f_1 \sin \tau + f_2 \cos \tau
  \right)\,, \\
 \end{aligned}
 \end{equation}
where the parameter $\varepsilon \propto a \propto \sqrt{q}$, which depends on the flux $q_0$ of
the plasma wave breaking at the critical point, does not exceed 1, and the functions
$f_{1,2}(\eta)$ are determined by the well-understood linear structure of the field, whose explicit
form can be various, depending on the density profile and the thermal motion of the electrons in
the plasma. In cold plasma with a linear density profile, we have
\begin{equation}\label{pl-cold}
 \begin{aligned}
  & f_1 = \left(1+ \eta^2 \right)^{-1} \,, \qquad
    f_2 = \eta \left(1+ \eta^2 \right)^{-1} \,. \\
 \end{aligned}
 \end{equation}
When a weak thermal motion of the electrons is taken into account, relations (\ref{pl-cold}) must
be modified:
 \begin{equation}\label{pl-hot}
 \begin{aligned}
  & f_1 = \int\limits_{0}^{\infty} d \xi \cos(\eta \xi + \xi^3/3) \,,
  \qquad
   f_2 = \int\limits_{0}^{\infty} d \xi \sin(\eta \xi + \xi^3/3) \,. \\
 \end{aligned}
 \end{equation}
Solution (\ref{solution-el}) is an exact solution of Eqns (\ref{twoeq}) for $\omega_L = \omega$.
The $x$ and $\eta$ variables in relations (\ref{solution-el}) -- %, (\ref{pl-cold}) and
(\ref{pl-hot}), in view of the normalization by the width of the plasma resonance $\Delta$, are
dimensionless quantities. The equations for the remaining four normalized quantities (the electric
field $E_y$, the magnetic field $B_z$, the $y$-component $V_y$ of the velocity of the electrons,
and the density $n$) are given by
 \begin{equation}
  \label{nonpotential}
 \begin{aligned} &
  \partial_{\eta} E_y  = - (\omega \Delta /c) \sin \theta \partial_{\tau} E_x
  \,, \quad
  \omega \partial_{\tau} V_y = E_y  \ ,   \\ &
  \partial_{\eta} B_z = (V_x/c) \partial_{\eta} E_y
                       - (V_y/c) \partial_{\eta} E_x\,, \quad
  n \thickapprox \omega^2 (1/\partial_{\eta} x) \,. \\
 \end{aligned}
 \end{equation}
Integration of Eqns (\ref{nonpotential}) is elementary. Formulas (\ref{solution-el}) and
(\ref{nonpotential}) present the required solution of the BVP. Discarding strongly nonlinear
effects, we can use (\ref{solution-el}) and (\ref{nonpotential}) to obtain results from the theory
of generation of arbitrary-order harmonics in cold \cite{sil-fp-80} and hot \cite{tro-diss}
inhomogeneous plasma (if we respectively use formulas (\ref{pl-cold}) and (\ref{pl-hot}) for
$f_{1,2}$). Taking strong nonlinearities (the influence of higher harmonics on the lower ones) into
account significantly changes the dependence of the coefficient of the transformation into
harmonics emitted by the plasma on the density of the electromagnetic radiation flux falling on the
plasma \cite{kov-tmf-89,kov-fp-89a} and the temperature of the plasma \cite{kov-fp-89b,kov-qe-89}.
\par

Result (\ref{solution-el}), (\ref{nonpotential}) of solving the BVP for the six original equations
takes both the boundary condition and the strongest nonlinearity into account, and is exact in the
same measure in which the group symmetry of Eqns (\ref{twoeq}) reflects the symmetry of the full
system of six original equations under the above assumptions. The approximate nature of the group
with infinitesimal operator (\ref{rg-el-gen}) so obtained relative to the group (\ref{group-el})
inducing it is determined by the inhomogeneity of the plasma (we recall that in the derivation of
operator (\ref{group-el}), we imposed no assumptions on Eqns (\ref{twoeq}) concerning the
inhomogeneity pattern of the plasma density). This is similar to the situation in quantum field
theory: the exact group property of a solution is used for a progressive improvement of the system
of its approximation characteristics, where the next approximation improves the previous one
without destroying it. From the standpoint of the RG symmetry, this means that operator
(\ref{rg-el-gen}) can be refined by accounting for the small parameters of the problem used in
passing to Eqns (\ref{twoeq}). We say in this case that the symmetry of system (\ref{twoeq}) is
\textit{inherited} by a more general system of equations. An example of an RG symmetry for
(\ref{twoeq}) with the corrections due to the inhomogeneity of the plasma taken into account is
presented in \cite{kov-tmf-89}.

\subsection{Renormalization-group symmetries in problems of gaseous and quasi-Chaplygin media} \label{rg-kcs}

The situation where the existence of an infinite-dimensional point transformation group ensures the
construction of an RG symmetry, as in the examples of BVPs for the Hopf equations and Eqns
(\ref{twoeq}), is not universal. Below, we present an example of a BVP in which the symmetry group
for the original manifold (the system of differential equations) is infinite dimensional, but the
construction of the RG symmetries requires using higher symmetries (which are also called
Lie-B\"acklund symmetries \cite{ibr-bk-83}) instead of a point transformation group. \par

We consider the BVP for a system of two nonlinear first-order partial differential equations for
functions $v$ and $n > 0$:
 \begin{equation}
 \label{kcs}
  \begin{aligned} &
 \partial_t v + v\partial_x v = \alpha \varphi (n) \partial_x n, \quad
 \partial_t n + v \partial_x n + n \partial_x v = 0\,, \\
 & v(0,x)=\alpha W(x) \,, \quad n(0,x)=N(x) \,,
 \end{aligned}
 \end{equation}
with constant $\alpha $ and a nonlinearity function $\varphi$ depending only on $n$. Depending on
the sign of $\alpha \varphi (n)$, these equations are of either the hyperbolic ( $\alpha \varphi
(n) < 0 $) or the elliptic ($\alpha \varphi (n) > 0 $) type. In the first case, (\ref{kcs})
corresponds to the standard equations of gas dynamics for one-dimensional planar isoentropic motion
of gas with the density $n$ and velocity $v$. The second case relates to equations of
quasi-Chaplygin media.\footnote{The term 'quasi-Chaplygin media' is used in the discussion of
nonlinear phenomena developing in accordance with the mathematical scenario for the Chaplygin gas,
i.e., the gas with a negative adiabatic exponent. At first glance, such a model looks like the
standard model of gas dynamics, but it corresponds to the negative first derivative of the
'pressure' with respect to the 'density.' A characteristic feature of quasi-Chaplygin media is a
universal mathematical form of various nonlinear effects accompanying the development of an
instability.} \par

Because Eqns (\ref{kcs}) are linear in the hodograph variables $\tau = nt$ and $\chi = x-vt$,
\begin{equation}
 \partial_v \tau - \psi(n)\, \partial_n \chi = 0\,, \quad
 \partial_v \chi + \partial_n \tau = 0 \,, \quad \psi = n/\alpha \varphi \, ,
 \label{god1}
 \end{equation}
there exists an infinitesimal operator of an infinite-dimensional subgroup $X_{\infty} = \xi^{1}
\partial_{\tau} +  \xi^{2} \partial_{\chi}$, whose coordinates $\xi^{1} $ and $\xi^{2}$ are defined
by arbitrary solutions of the partial differential equations transformed into (\ref{god1}) by the
substitution $\tau \to \xi^{1}$, $\chi \to \xi^{2}$. This means that we can formally construct an
RG symmetry by restricting an infinite-dimensional point transformation group, but this requires
knowing the solution of (\ref{god1}) for arbitrary boundary data; in fact, such a procedure is
equivalent to solving the original BVP. \par

Another approach to the construction of RG symmetries for the problem under consideration is the
use of a higher (Lie-B\"acklund \cite{ibr-dan-76}) symmetry group. By contrast to point
transformation groups with generators of form (\ref{oper-def}), Lie-B\"acklund symmetries are
characterized by an infinitesimal operator with the coordinates depending on independent variables
$x$ and differential variables $u$ and the derivatives $u_{(1)} = \{\partial_{x^i} u^{\alpha} \}
\equiv \{u^{\alpha}_{i} \}$, $u_{(2)} = \{
\partial_{x^i x^j} u^{\alpha} \}\equiv \{u^{\alpha}_{ij} \},\ldots $, where $\alpha = 1,\ldots , m$; $ i,j =
1,\ldots , n$. The relation between these variables can be expressed by means of the total
differentiation operators $D_i$, as the following system of equalities:
 \begin{equation}
 \label{rel-def}
  u^{\alpha}_{i} = D_i \left( u^{\alpha} \right) \,, \
  u^{\alpha}_{ij} = D_j \left( u^{\alpha}_{i} \right) =
                    D_j D_i \left( u^{\alpha} \right)\,, \ldots ,  \
  D_i = \partial_{x^i} +
      u^{\alpha}_{i} \partial_{u^{\alpha}} +
      u^{\alpha}_{ij} \partial_{u^{\alpha}_{j}} +
      \ldots .
 \end{equation}
The Lie-Backlund group theory allows restricting to only canonical operators, which leave all the
independent variables invariant. This is important, for instance, in the analysis of symmetries of
integrodifferential equations and in the construction of the RG symmetries in problems involving
nonlocal equations. For BVP (\ref{kcs}) under consideration here, it is convenient to write the
Lie-B\"acklund group generators for the equations expressed in hodograph variables (\ref{god1}):
\begin{equation}
 X= \sum\limits_{i} c_i X_i \equiv  \sum\limits_{i} c_i
 \left( f_i \partial_{\tau} + g_i \partial_{\chi} \right) \,.
 \label{rg-liebk-kcs}
 \end{equation}
The coordinates $f_i$ and $g_i$ of generators (\ref{rg-liebk-kcs}), which are linear functions of
the differential variables \cite{kpu-mcm-97}, are connected by a system of recursion relations
\begin{equation}
  \begin{array}{c}
 L_k
 \end{array}
 \left(
 \begin{array}{c}
 f_j \\ g_j
 \end{array}
 \right)
 =
 \left(
 \begin{array}{c}
 f_{j+1} \\ g_{j+1}
 \end{array}
 \right) ,
 \label{rg-liebk-rec}
 \end{equation}
where the entries of the matrix recursion operators $L_k$ are linear functions of the operator
$D_n$ of total differentiation with respect to $n$. The number of operators $L_k$ depends on the
form of the nonlinearity function $\varphi (n)$; in the most typical case $\varphi(n)
=n(n+b)^\ell$, where $b,\ \ell=$ const, there are three operators: $k = 1, 2, 3$. The action of the
recursion operators on the coordinates $f_1=\tau$ and $g_1=\chi$ of the physically 'obvious'
dilation operator in the space of the hodograph variables $\tau = nt$ and $\chi = x-vt$ yields
three operators with coordinates $f_i$ and $g_i$ $(i = 2,3,4)$ linearly depending on the
derivatives $\tau_n = \partial_n \tau$ and $\chi_n = \partial_n \chi$; they are therefore
equivalent to infinitesimal operators of the point group. The action of the recursion operators
$L_1,\ldots , L_3$ on the first-order symmetries $f_i, g_i$ $(i = 2,3,4)$ generates five operators,
whose coordinates in the hodograph variables are linear functions of these variables and their
second-order derivatives. These are Lie-B\"acklund symmetries of the second order. Repeating this
procedure several times, we obtain $2s + 1$ symmetries of a fixed order $s$ \cite{kpu-mcm-97}.
\par

The infinite system of operators (\ref{rg-liebk-kcs}) (obtained at step \textbf{(II)} of the RG
algorithm) for Eqns (\ref{god1}) (treated as the $\cal{RM}$ manifold) enables constructing the
operators of RG symmetries and finding the corresponding RG-invariant solutions. The reduction of
the Lie-B\"acklund group (step \textbf{(III)} of the RG algorithm) reduces to the verification of
the invariance conditions $f = 0$ and $g = 0$ (similar to (\ref{invarcond-hopf}), but generalized
to the case of Lie-B\"acklund symmetries) for a concrete solution of the BVP, where the functions
$f$ and $g$ are arbitrary linear combinations of some coordinates $f_i$ and $g_i$ of the canonical
operators of the group and are chosen so as to satisfy the prescribed boundary conditions at $t=0$.
As examples, we give the values of the coordinates of two second-order operators of the Lie -
Backlund RG symmetry.

\noindent
 {\textit{Example 1.}}
  \begin{equation}
  \label{rgs1_soliton}
  \begin{aligned}
  & f = 2n(1-n)\tau_{nn}- n\tau_n - 2nv(\chi_n + n \chi_{nn})+
    n v^2 \tau_{nn} /2  \,,  \\
  & g = 2n(1-n)\chi_{nn}  +(2-3n)\chi_n +
     v \left( 2n\tau_{nn} + \tau_n \right)
     + (v^2 /2)\left( n \chi_{nn} + \chi_n \right) \,.
           \\
 \end{aligned}
 \end{equation}
\par

\noindent {\textit{Example 2.}}
 \begin{equation}
 \label{rgs_slab}
 \begin{aligned}
 & f = - n^2 \ln n \tau_{nn} - (n/2) \tau_n + \tau /2
   + v (n^3\chi_{nn} + (3/2) n^2 \chi_n ) \,, \\
 & g =  - n^2 \ln n  \chi_{nn} + (n/2)(1+4 \ln n )\chi_n + \chi /2  +
      v \left( n \tau_{nn} + \tau_n /2 \right) \,.
 \end{aligned}
 \end{equation}
The operator $R$ with coordinates (\ref{rgs1_soliton}) corresponds to the solution of the BVP for
Eqns~(\ref{kcs}) with $\alpha=1$, $\varphi (n) = 1$ for $W(x) = 0$ and $N(x) = \cosh^{-2}(x)$, and
the operator $R$ with coordinates (\ref{rgs_slab}) corresponds to the solution of the BVP for Eqns
(\ref{kcs}) with $\alpha=-1$, $\varphi (n) = 1/n$ for $W(x)=0$ and $N(x)= \exp (-x^2)$. To solve
the BVP using RG symmetries (\ref{rgs1_soliton}) and (\ref{rgs_slab}), we must add the invariance
condition $f=g=0$ to the basic $\cal{RM}$ and solve the resulting system of equations (step
\textbf{(IV)} of the RG algorithm). \par

For RG symmetry (\ref{rgs1_soliton}), a solution exists on a finite interval $0 \leqslant t
\leqslant t_{sing} $, until a singularity occurs on the axis $x=0$ at $t=t_{sing}=1/2$, when
$\partial_x v (t_{sing},0) \to \infty$ and the value of $n$ remains finite, $n (t_{sing},0) = 2 $:
\begin{equation}
 v = - 2 nt \tanh (x-vt)\,, \quad  n^2t^2= n\cosh^2(x-vt)-1\,.
 \label{res1_soliton} \end{equation}
From the physical standpoint, solution (\ref{res1_soliton}), which was previously obtained in
\cite{ahm-jetf-66}, describes the evolution of a planar light beam in a medium with a cubic
nonlinearity (a quasi-Chaplygin medium) for the boundary condition $N(x)= \cosh^{-2}(x)$. The
quantities $n$ and $v$ define the intensity and the eikonal derivative of the beam. \par

For RG symmetry (\ref{rgs_slab}), the solution describes a monotonic evolution (decrease) with time
$t$ of the density $n \geqslant 0$, while the particle velocity continues to be linearly dependent
on the coordinate:
\begin{equation}
 v = x \sqrt{2} q {\rm e}^{-q^2/2}\, , \quad
 n = {\rm e}^{-q^2/2} \exp \left( - x^2 {\rm e}^{-q^2}  \right)  \,,
 \quad t=(\sqrt{\pi}/2) {\rm{erfi}} \left(
  q/\sqrt{2} \right)  \, .
 \label{res2_murakami} \end{equation}
Solution (\ref{res2_murakami}), which was discussed in \cite{mur-pop-05}, describes an expanding
plasma layer with the initial density distribution $N(x)=\exp (-x^2)$. \par

These two examples demonstrate that by using the Lie-B\"acklund RG symmetry, we achieve the same
goals as with point RG symmetries: we can give an adequate description of the structure of the
solution in the presence of a singularity or can find its asymptotic behavior. Although we found RG
symmetries (\ref{rgs_slab}) and (\ref{res1_soliton}) for the already known solutions, the RG
approach reveals the group structure of these solutions. Previously, to obtain these results, the
authors imposed some \textit{a priori} assumptions about the structure of the solution. In
\cite{kov-tmf-97}, the reader can find an example of the solution of a BVP with the help of
Lie-Backlund RG symmetries for (\ref{kcs}) with the initial condition of a more complex type, not
representable in terms of elementary functions, when the intensity distribution of the light beam
at the boundary has the form of a smoothed step function. \par

\subsection{Approximate renormalization-group symmetries in problems of quasi-Chaplygin media}{\label{rg-apr-ngo}}

Constructing an RG symmetry on the basis of higher symmetries is justified if the equations
defining an RG-invariant solution can be investigated analytically. The complexity of differential
equations usually increases with their order. Hence, the use of higher-order Lie-B\"acklund
symmetries in the invariance conditions of the RG symmetry can often limit the potential for
applications of such symmetries in the case of arbitrary boundary data. On the other hand, a
restriction on the order of the allowed symmetries narrows the variety of approaches to the
construction of RG symmetries for arbitrary boundary data. For instance, for BVP (\ref{kcs}), the
symmetry group of the original manifold (\ref{god1}) allows only $2s + 1$ symmetries of a fixed
order $s$, which for small $s$ can be insufficient for the construction of the RG symmetry for
arbitrary $N(x)$. For the extension of the symmetry group of the original manifold, we must use the
technique of approximate symmetries \cite{bgi-msb-88}.

The central idea here is the use of natural smallness parameters (which we distinguish from the
parameter with respect to which we construct the PT approximation to be used in RG
transformations), which are involved in some form in most physical problems and which enter the
equations as coefficients. For instance, the coefficient $\alpha$ of the nonlinearity function
$\varphi (n)$ in (\ref{kcs}) is such a parameter. The presence of natural small parameters allows
expressing the required symmetry as a power series in these parameters and taking finitely many
terms of this series. If we discard the small parameters altogether, then the equations defining
the $\cal{RM}$ are simpler than the original equations and allow a wider transformation group, and
hence there can be more approaches to the construction of the RG symmetry for arbitrary boundary
data. An essential point here is the possibility to successively account for corrections to the
obtained RG symmetry for the system of differential equations of the simplified manifold: when this
can be done, we say that we have constructed a symmetry inherited in a given order in the small
parameter. \par

We demonstrate how approximations to the RG symmetry for BVP (\ref{kcs}) can be constructed for
small $\alpha \ll 1$. Setting $w=v/\alpha$, we write system of equations (\ref{god1}) as
\begin{equation}
 \partial_w \tau - (n / \varphi(n))\, \partial_n \chi = 0\,, \quad
 \partial_w \chi + \alpha \partial_n \tau = 0 \,.
 \label{god2}
 \end{equation}
As $\alpha \to 0$, dropping the second term in the second equation yields a simpler subsystem of
differential equations, which is an approximation to the original manifold $\cal{RM}$. By contrast
to the symmetries of Eqns (\ref{god1}), which allow only a finite-dimensional Lie-B\"acklund
symmetry group of a given order, Eqns (\ref{god2}) with $\alpha=0$ have an infinite-dimensional
symmetry, which is consistent with the perturbation theory for the BVP with arbitrary boundary
data. Hence, we seek RG symmetries by combining symmetries of the 'zeroth' approximation to the
equations (i.e., of Eqns (\ref{god2}) with $\alpha=0$) and corrections to them in powers of
$\alpha$. We represent the coordinates $f$ and $g$ of the canonical operator of the group for
(\ref{god2}) as a power series in $\alpha$:
 \begin{equation}
 X=f \partial_\tau + g \partial_\chi\,, \quad  f=\sum\limits_{i=0}^{\infty} \alpha^i f^i \,; \quad
   g=\sum\limits_{i=0}^{\infty} \alpha^i g^i\,.
 \label{ser1} \end{equation}
Using the techniques of modern group analysis \cite{ibr-spr-94} that generalize Lie's
algorithm to higher symmetries, we obtain a system of recursive relations for the $f^i$ and $g^i$:
 \begin{equation}
 \begin{aligned}
 f^i & = F^i + \int {\rm d} w \left\{ (1-\delta_{i,0}) Z
 f^{i-1} + \frac{n}{\varphi} Y g^i \right\}, \\
 g^i & = G^i + (1-\delta_{i,0}) \int {\rm d} w
 \left\{Z g^{i-1} -  Y f^{i-1} \right\}\,, \\
 \end{aligned}
 \label{rec1}
 \end{equation}
where
 \begin{equation}
 \begin{aligned}
 Y & = \partial_n + \sum\limits_{s=0}^{\infty}\left (
 \tau_{s+1} \partial_{\tau_s} + \chi_{s+1}\partial_{\chi_s} \right)
 \,, \quad Z=\sum\limits_{s=0}^{\infty}  \tau_{s+1}
 \partial_{\chi_s} \,,   \quad w = v / \alpha \, , \\
 \tau_s & =  \frac{ \partial^s \tau}{ \partial n^s }\,, \quad
 \chi_s = \frac{ \partial^s \chi }{\partial n^s }\,, \quad
 \tilde{\tau}_s = \tau_s - w \sum_{p=0}^{s} {s \choose p}
 \frac{\partial^p ( n / \varphi)}{\partial n^p } \chi_{s-p+1}
 \,,\\
 \end{aligned}
 \label{rec2} \end{equation}
$F^i(n,\chi_s,\tilde{\tau}_s)$ and $G^i(n,\chi_s,\tilde{\tau}_s)$ are arbitrary functions; the
integrands are expressed in terms of $\tilde{\tau}_s,\, \chi_s, \, n,$ and $w$. It is an immediate
consequence of (\ref{rec1}) that for small~$\alpha$, the symmetry of the equations of the 'zeroth'
approximation is inherited by system (\ref{god2}) in any finite order in ~$\alpha$: the corrections
do not destroy the symmetry $f^0, g^0$ of the 'zeroth' approximation. The form of the inherited
symmetry (i.e., the expressions for $f$ and $g$) is fully determined by relations (\ref{rec1}): it
can be a point symmetry or a Lie-B\"acklund symmetry. \par

Because the dependence of the functions $F^i$ and $G^i$ on their arguments can be arbitrary, we can
construct RG symmetries for the BVP with arbitrary boundary data: the restriction of approximate
group (\ref{rec1}) on solutions of the BVP (step \textbf{(III)} of the RG algorithm) is performed,
similarly to the case of the exact Lie-Backhand RG symmetry in subsection~\ref{rg-kcs}, by
verifying the condition $f=g=0$ for a concrete solution of the BVP. Here, we choose the functions
$F^i$ and $G^i$ so as to satisfy the prescribed boundary conditions at $t = 0$. In particular, for
BVP (\ref{kcs}) with $W(x)=0$, we can set $F^i$ and $G^i$ $(i\geqslant 1)$ equal to zero and can
ensure the boundary conditions by selecting $F^0$ and $G^0$. \par

We now present two examples of RG symmetries constructed with the use of relations (\ref{rec1}).
The first example is related to the BVP for Eqns (\ref{kcs}) with $\varphi (n) = 1$ for $W(x) = 0$
and $N(x)= \cosh^{-2}(x)$. With these conditions, we can take the following functions $f^0$ and
$g^0$:
 \begin{equation}
  \label{f0g0}
  \begin{aligned}
  & f^0 = 2n(1-n)\tau_2- n\tau_1 - 2n w (\chi_1 + n \chi_2 ) \,,  \quad
  & g^0 = 2n(1-n)\chi_2  +(2-3n)\chi_1 \,.
 \end{aligned}
 \end{equation}
Substituting $f^0$ and $g^0$ in (\ref{rec1}), we find the next terms of series (\ref{ser1}), the
functions $f^1$ and $g^1$:
\begin{equation}
  \label{f1g1}
  \begin{aligned}
  & f^1 = n w^2 \tau_2 /2  \,,  \quad
    g^1 = w \left( 2n\tau_2 + \tau_1 \right)
     + (w^2 /2)\left( n \chi_2 + \chi_1 \right) \,,
 \end{aligned}
 \end{equation}
and the substitution of $f^1$  and $g^1$ in (\ref{rec1}) gives zero values for all $f^i$ and $g^i$
with $i \geqslant 2$. This means that the RG symmetry can be expressed in this case by binomials
$f=f^0 + \alpha f^1$, $g=g^0 + \alpha g^1$, that is, infinite series (\ref{ser1}) terminate and
turn into finite sums, and the binomial expressions for the RG symmetry are exact and hold for
arbitrary values of $\alpha$. In particular, setting $\alpha=1$, we arrive at relations
(\ref{rgs1_soliton}).
\par

For arbitrary boundary data, the infinite series in (\ref{ser1}) do not automatically terminate,
and taking only finitely many terms of the series means that the RG symmetry constructed with the
use of (\ref{ser1}) and (\ref{rec1}) is approximate in the sense described in \cite{bgi-vin-89}.
The second example corresponds to an approximate RG symmetry for BVP (\ref{kcs}) with $\varphi (n)
= 1$ for $W(x)=0$ and $N(x)= \exp (-x^2)$:
 \begin{equation}
  f = 1+ 2n\chi \chi_n +\alpha \left( -2 \tau \tau_n +\tau^2 /n \right) \,,
  \quad  g =  -2\alpha \left(\tau \chi_n + \chi \tau_n  \right) \,.
 \label{gauss1}
\end{equation}
Here, we omit all the contributions to $f$ and $g$ proportional to the higher powers $\alpha^i$
with $i \geqslant 2$. \par

The above constructions of RG symmetries can be easily generalized to the case where the group
transformations involve the parameter $\alpha$ in addition to the 'natural' variables of the
problem. In this case, the set of possible RG symmetries is usually larger. For example, we note
the approximate RG symmetry for the same BVP as in the second example, but, in contrast to
(\ref{gauss1}), containing derivatives with respect to the parameter $\alpha$:
 \begin{equation}
  f =  2n(\tau \chi_n + \tau_n\chi) + 2 \alpha \chi \tau_{\alpha} \,,\quad
  g = 1 + 2n\chi \chi_n + 2 \alpha \left(\chi \chi_{\alpha} - \tau \tau_n \right)\,.
 \label{gauss2}
\end{equation}
Unlike exact RG symmetries, which allow finding an exact solution of the BVP for any RG generator
chosen, approximate symmetries yield a solution of the BVP depending essentially on the form of the
RG symmetry operator, as can be seen, for instance, from the use of generators (\ref{gauss1}) and
(\ref{gauss2}) (see \cite{kov-tmf-97}). \par

The use of several approximate analytic solutions or the comparison of the solution obtained on the
basis of the exact RG symmetry (and used as a test) with the solution obtained on the basis of an
approximate RG-symmetry allows evaluating the accuracy of the corresponding approximate
RG-invariant solution \cite{kov-nd-00}.
\par

For finding approximate RG symmetries in a physical problem, we can use not one but several small
parameters. This is the case, for instance, in the construction of the RG symmetry for a BVP for
the system of equations of a light beam in a nonlinear medium, which can be regarded as a natural
generalization of (\ref{kcs}):
 \begin{equation}
 \label{basic}
 \begin{aligned} &
 v_{t}  +  v v_{x}  -  \alpha \varphi(n) n_{x}
   - \beta \partial_{x}\left(\left(x^{-\nu}/\sqrt{n \,} \right)
    \partial_{x}\left( x^{\nu }\partial_{x} \left( \sqrt{n\,}
     \right) \right) \right) = 0 \,,
 \\ &
 n_{t}  +  n v_{x} +  v n_{x}  +\nu(nv/x) = 0\,, \quad v(0,x)= V(x) \, , \quad
  n(0,x)=N(x) \, .\\
 \end{aligned}
 \end{equation}
The parameters $\alpha$ and $\beta$ determine the contribution of the nonlinear refraction and
diffraction processes; $\nu = 0$ for a planar (2-dimensional) wave beam and $\nu=1$ for a
3-dimensional (axially symmetric) wave beam. \par

The construction of an RG symmetry for BVP (\ref{basic}) proceeds in accordance with a scheme
similar to the one used before: the coordinates $f$ and $g$ of the canonical operator for the
manifold $\cal{RM}$ defined by Eqns (\ref{basic}) can be represented as double power series in the
nonlinearity parameter $\alpha$ and the diffraction parameter $\beta$:
 \begin{equation}
  X = f\partial_{v} + g\partial_{n}\, , \quad
   f=\sum\limits_{i,j=0}^{\infty} \alpha^i \beta^j f^{(i,j)} \,, \quad
   g=\sum\limits_{i,j=0}^{\infty} \alpha^i \beta^j g^{(i,j)} \,.
 \label{fg-gen} \end{equation}
The standard techniques of group analysis are used for the calculation of the coefficients
$f^{(i,j)}$ and $g^{(i,j)}$. Restricting ourself to finitely many terms of series (\ref{fg-gen}),
we arrive in the general case at an approximate symmetry, which after the restriction procedure
gives the required RG symmetry. As an example \cite{kov-tmf-99}, we present explicit expressions
for the coordinates $f$ and $g$ of the infinitesimal RG symmetry operator for BVP (\ref{basic}) in
the case of a collimated cylindrical $(\nu=1)$ beam in a medium with cubic nonlinearity $(\varphi =
1)$:
 \begin{equation}
 \label{fg-cyl-cub}
 \begin{aligned} &
 f  = D_x \left\{ S - \left( \alpha n +
 \frac{\beta}{x\sqrt{n}} D_x \left( xD_{x}\sqrt{n\,}\right)\right)
   \right\}  \,, \quad  g  = \frac{1}{x} D_x \left\{ (xn) \left[v- t S_{\chi} \right]\right\}  \,, \\
 \end{aligned}
 \end{equation}
where the function $S$ depends on $\chi=x-vt$:
 \begin{equation}
 \label{S} S(\chi) = \alpha N(\chi) +
 \frac{\beta}{\chi\sqrt{N(\chi)}} \,
 \partial_{\chi} \left( \chi \partial_{\chi} \left( \sqrt{N(\chi)}\right)
 \right) \,. \end{equation}
The canonical RG operator with coordinates (\ref{fg-cyl-cub}) is equivalent to the following
operator of a point RG symmetry:
 \begin{equation}
 \label{rgsym}
 \begin{aligned} &
 R = \left[   1 + t^2 S_{\chi\chi} \right]    \partial_{t}
     + S_{\chi} \partial_{v} + \left[ t S_{\chi} + v t^2 S_{\chi\chi} \right] \partial_{x}
 - nt \left[ \left(1+\frac{vt}{x} \right) S_{\chi\chi}
   + \frac{1}{x} S_{\chi} \right] \partial_{n} \,, \\
 \end{aligned}
 \end{equation}
which allows easily finding finite group transformations (step \textbf{(IV)} of the RG algorithm)
relating the values of $n$ (the beam intensity) and $v$ (the eikonal derivative) for $t>0$ to
similar quantities at the boundary $t=0$ of the nonlinear medium, i.e., constructing the required
solution of BVP (\ref{basic}) \cite{kov-tmf-99}. \par

In the derivation of (\ref{S}) and (\ref{rgsym}), we considered contributions of the form
$f^0\equiv f^{(0,0)}$ and $g^0\equiv g^{(0,0)}$ in (\ref{fg-gen}), that is, contributions
independent of $\alpha$ and $\beta$, and also contributions linear in these parameters $f^1\equiv
\alpha f^{(1,0)} + \beta f^{(0,1)}$ and $g^1\equiv \alpha g^{(1,0)}+\beta g^{(0,1)}$. Dropping the
terms proportional to $O\left(\alpha ^2, \beta ^2, \alpha \beta \right)$ means that symmetry
(\ref{rgsym}) is approximate with respect to these parameters. Of course, similarly to BVPs for
equations of quasi-Chaplygin media, there exist distributions $N(x)$ for which series
(\ref{fg-gen}) terminate and become finite sums. Such a situation corresponds to the exact RG
symmetry in (\ref{fg-cyl-cub}) rather than an approximate one and to an exact solution of the BVP
for arbitrary values of the parameters $\alpha$ and $\beta$. In particular, symmetry (\ref{fg-gen})
is exact when $S(x)$ is a binomial: $S(x) = s_0 + s_2 x^2/2$. This form of $S(x)$ corresponds to a
particular dependence on the $x$ variable of the beam intensity $N$ at the interface. For instance,
for $s_2=0$ and $s_0>0$, Eqn (\ref{rgsym}) yields a solution of the BVP that describes a Townes'
self-channeling beam \cite{tow-prl-64}; other exact localized solutions of the BVP for $s_2 \neq 0$
decreasing as $x \to \infty$ were discussed in \cite{kov-pre-00}. \par

In the general case, $S(x)$ is not a binomial and the use of RG symmetry (\ref{rgsym}) yields an
approximate analytic solution of the BVP, studied in detail in \cite{kov-tmf-99,kov-pre-00} for a
Gaussian beam with $N=\exp (-x^2)$. This solution of the BVP allows tracing the evolution of the
Gaussian beam as the distance from the interface increases, up to a singularity occurring in the
solution; the singularity has the 2-dimensional structure: both the beam intensity $n$ and the
derivatives $v_x$ and $n_x$ become infinite at the point $t_{sing}^{Gauss} = 1/\sqrt{ 2 (\alpha -
\beta)}$ for $\alpha > \beta$. A thorough discussion of this analytic solution and its comparison
with the results of other approaches (aberration-free approximation \cite{ahm-jetf-66} and the
method of moments \cite{tal-izv-71,tal-bk-97}) were carried out in \cite{kov-pre-00}.

\section{Renormalization-group symmetries in nonlocal problems of mathematical physics \label{rgs-nlc}}

The implementation of the RG algorithm in problems of mathematical physics involving nonlocal
(integral or integrodifferential) relations depends on the form of this nonlocality. On the one
hand, the original system of equations can be based on nonlocal relations, as, for example, in the
kinetic plasma theory, according to which the relation between the current density and the charge
density in a medium and the distribution function of the plasma particles in the Vlasov- Maxwell
equations with a self-consistent field is nonlocal. The application of the RG algorithm to such
nonlocal problems of mathematical physics proceeds in accordance with the general scheme in
Section~\ref{illustration}; the difference is in the methods of the calculation of symmetries for
nonlocal objects (see \cite{kov-nd-00} and the references therein). We note that in the case of
problems described by complicated equations, as in transfer theory (the Boltzmann
integrodifferential equation), only some components of the solution or its integral characteristics
can have a relatively simple symmetry. For instance, in the simplest planar one-velocity transfer
problem, the RG invariance is a property of the asymptotic form as $x \to \infty\,$ of the 'density
of particles going inside the medium' $n_{+}(x)$, which does not feature in the Boltzmann
equation.\footnote{But it can be represented as the integral $ \int\limits_{0}^{1}
n(x,\vartheta)\,{\rm{d}} \cos\vartheta\,$ of the solution $n(x,\vartheta)\,$ of the one-velocity
kinetic equation.} \par

On the other hand, interesting from the physical standpoint can often be not the solution itself
over the entire range of its arguments and parameters but some integral characteristic, a
functional of the solution. For instance, this characteristic can be the result of averaging
(integrating) with respect to some independent variable or of passing to another integral (e.g.,
Fourier) representation. In this case, we can use the RG algorithm to improve the functional of the
approximate solution rather than to improve the particular solution and the subsequent calculation
of the corresponding integral characteristic. \par

We now present examples of the implementation of the RG algorithm in problems of mathematical
physics involving nonlocal relations, which provide illustrations to both possible cases.

\subsection{Renormalization-group symmetries for functionals of solutions\label{rg-functionals}}

We consider some BVP for local equations and assume that we are interested in an integral
characteristic of the solution, given by a linear functional of this solution:
 \begin{equation} \label{rgmnonloc}
 \begin{aligned} &
 J(u) = \int {\cal F}(u(z)) {\rm{d}} z\,.
 \end{aligned}
 \end{equation}
We assume that for a particular solution $u$ of this boundary value problem, the RG algorithm has
been used to find an RG symmetry with a generator $R$. Instead of the RG transformation group of
the solution itself, we are interested in the RG transformation group of integral characteristic
(\ref{rgmnonloc}). To find an infinitesimal generator of the group, we extend the action of the RG
symmetry operator $R$ to nonlocal variable (\ref{rgmnonloc}). For this, we represent the operator
in the canonical form, that is, make the substitution $R \to Y={\varkappa} \partial_u$, and extend
the operator to the nonlocal variable $J$:
 \begin{equation} \label{prolong-can}
 Y + {\varkappa}^J \partial_{J} \equiv {\varkappa} \partial_u
                      + {\varkappa}^J \partial_{J} \,.
 \end{equation}
The ${\varkappa}^J$ variable is related to ${\varkappa}$ by means of an integral relation
\cite{ibr-nd-01} (for brevity, we write only one argument of the coordinate of the generator, the
one with respect to which the integration is performed):
 \begin{equation}
 \label{canonic3}
 \varkappa^J = \int \frac{\delta J(u)}{\delta u (z)}\, \,
 \varkappa(z)\,    {\rm{d}} z \equiv
 \int \frac{\delta {\cal F}(u(z^{\prime}))}{\delta u (z)} \,\,
 \varkappa(z)\, {\rm{d}} z \, {\rm{d}} z^{\prime} =
  \int {\cal F}_u \,\, \varkappa(z)\,  {\rm{d}} z \,.
 \end{equation}
Considering operator (\ref{canonic3}) in the narrowed space of the variables defining the
functional, we obtain the required infinitesimal RG symmetry operator for integral characteristic
(\ref{rgmnonloc}). \par

\subsubsection{The RG symmetry of functionals in the Hopf equation}

To demonstrate how formulas (\ref{prolong-can}) and (\ref{canonic3}) actually work for functionals
of solutions of BVPs, we start with our example of the BVP for the Hopf equation. The algebra of RG
symmetries of this problem is generated by the three operators in (\ref{rg-hopf}). We consider the
case where we are interested not in the full solution to BVP (\ref{solution-hopf}) for all values
of its arguments and parameters but only in the value at some point of some characteristic, which
can be defined by a linear functional of form (\ref{rgmnonloc}). For instance, we can be interested
in the value of the first spatial derivative at $x=0$:
\begin{equation} \label{functional}
 \partial_x u(t,x)_{\vert x=0} \equiv u^0_x = - \int\limits_{-\infty}^{+\infty}
 \textrm{d}x \, \delta^{\prime} (x) u(t,x)\,.
\end{equation}
Using perturbation theory (\ref{hopf-pt1}) in the right-hand side of (\ref{functional}) yields the
behavior of $u^0_x$ for small $t \ll 1$:
\begin{equation} \label{hopf-pt2}
  (u^0_x)_{PT} = U_x^0 - \epsilon t \left[(U_x^0)^2 + U(0)U_{xx}^0 \right]
  + O \left( t^2 \right) , \ U_x^0 \equiv \partial_x U\big\vert_{x=0} ,
  \ U_{xx}^0 \equiv \partial_{xx} U\big\vert_{x=0} \,.
\end{equation}
To correct the asymptotic behavior of the functional of the solution, which is distorted by
perturbation theory (\ref{hopf-pt2}), we can use the RG symmetry for (\ref{functional}). As in the
derivation of solution (\ref{solution-hopf}), we use the last generator in (\ref{rg-hopf}) in the
simplest form, with $\epsilon \psi^4 = 1$, and write it in the space of variables
$\{t,\epsilon,u^0_x\}$. For instance, for $U=x\,,$ this operator is
\begin{equation} \label{rg-hopf-example1}
 R =\partial_{\epsilon}-t (u^0_x)^2\partial_{u^0_x}\,.
\end{equation}
Information about the behavior of the function $u^0_x=1/\left(1+\epsilon t \right)\,,$ in the
entire range of the $t$ variable, including its asymptotic behavior as $t \to \infty$, can be
obtained either with the help of (finite) transformations of the group with generator
(\ref{rg-hopf-example1}), which are similar to (\ref{trans-hopf}), or by using the obvious
invariant $J^0 = \epsilon t - 1/u^0_x$ of RG generator (\ref{rg-hopf-example1}), with the initial
condition $u^0_x(t=0)=1$. We emphasize that we obtain this result without explicitly finding
solution (\ref{solution-hopf}), but only using the RG symmetry. Our construction may look
cumbersome at first glance; however, in more complex problems, the solution is typically not known
explicitly, but the RG symmetry can be constructed (see, e.g.,~\cite{shk-mathphys-05}).

\subsubsection{The RG symmetry of functionals in quasi-Chaplygin media} One example of a more complicated
situation is the behavior of functionals of solutions of the BVP for the equations of
quasi-Chaplygin media (\ref{kcs}) and, more specifically, of the quantities $n(x)$ and $v(x)$ on
the axis $x=0$, up to the point where a singularity occurs. We claim that this phenomenon can be
investigated by applying the RG algorithm to two functionals of solutions of BVP (\ref{kcs}): the
density $n^{0}(t)\equiv n(t,0) $ and the derivative of the velocity ${W}^{0}(t) \equiv v_{x}(t,0)$
calculated on the axis of the beam and related to the solution by the formal equalities
 \begin{equation}\label{intensity-eikonal}
 n^{0}(t) = \int \textrm{d}x \, \delta (x) n(t,x)\,, \quad
 {W}^{0}(t) = \int \textrm{d}x \, \delta (x) v_x(t,x)\,.
 \end{equation}
The boundary conditions for functionals (\ref{intensity-eikonal}) can be written as
 \begin{equation}  \label{boundary-func}
  n^0(0)=1, \quad {W}^0(0)=0 \,.
  \end{equation}
Although conditions (64) give no information about the dependence of the density $n$ on the $x$
variable, such information is incorporated into the RG symmetry operator, whose explicit form is
determined by the density distribution $N(x)$ for $t=0$. We consider an example of the problem with
the planar geometry, with the 'soliton' profile $N(x)=\cosh^{-2}(x)$ of the density distribution at
the boundary, for the RG symmetry operator as in (\ref{rg-liebk-kcs}), (\ref{rgs1_soliton}).
Extending this operator to nonlocal variables (\ref{intensity-eikonal}), we obtain a simpler
operator in the space $\{t, \, n^0\}$ \cite{ksh-rg-05}:
 \begin{equation}
 \begin{aligned} &
 R=\varkappa^{n^0} \partial_{n^0} ,  \
  \varkappa^{n^0}=4-5n^0 -t  n^0_t + 2 ( n^0 -1) n^0 n^0_{tt} (n^0_t)^{-2}, \
  n^0_{t} \equiv \partial_t n^0, \  n^0_{tt} \equiv \partial_{tt} n^0  .
 \end{aligned}
 \label{rgs-pr-sol}
 \end{equation}
The RG invariance condition $\varkappa^{n^0}=0$ for operator (\ref{rgs-pr-sol}) leads to an
ordinary second-order differential equation for the function $n^0(t)$, which must be solved with
initial conditions (\ref{boundary-func}) and the additional condition $(n^0_t/
\sqrt{n^0-1})\big\vert_{t \to 0}=2$ for the first derivative, which follows from the original
equations (\ref{kcs}) for $x=0$. This solution $t = \sqrt{n^0-1}/ n^0$ reproduces the result
obtained in (\ref{res1_soliton}), but the method is simpler and solution (\ref{res1_soliton}) is
not explicitly required. \par

We note that the procedure of extending RG generators represented as infinitesimal operators of a
point group or a Lie-Backhand group is universal, and hence we have a common framework for the
description of the behavior of characteristics of solutions of BVPs (\ref{hopf}) and (\ref{kcs})
alike, if we use the reduced description in terms of functionals of solutions.
\par

\subsection{Renormalization-group symmetry in the problem of an expanding plasma bunch}
{\label{plasma-dyn}}

We now consider the construction of the RG symmetry in the problem where nonlocal relations are
involved in the definition of the basic manifold. We discuss the problem of an expanding plasma
bunch in the quasineutral approximation \cite{kov-jetf-02}. In this approximation, the dynamics of
plasma particles in the planar geometry are determined by solutions of kinetic equations for the
distribution functions $f^\alpha (t,x,v)$ of the different kinds of particles (electrons and ions):
\begin{equation}
 \partial_t f^\alpha  + v  \partial_x f^\alpha + (e_\alpha/m_\alpha)
 E(t,x)  \partial_v f^\alpha  = 0
 \label{bunch-kin}
 \end{equation}
with additionally imposed nonlocal constraints arising from the conditions of the vanishing current
and charge density (the quasineutrality conditions):
  \begin{equation}
 \int \textrm{d} v \, \sum\limits_{\alpha} e_\alpha f^\alpha = 0\,, \quad
 \int \textrm{d} v \,v \, \sum\limits_{\alpha} e_\alpha f^\alpha  =  0\,.
 \label{quasineutral}
 \end{equation}
The electric field is here expressed in terms of moments of the distribution functions:
 \begin{equation}
 \label{quasineutral2}
 E(t,x) = \left( \int \textrm{d}v \,v^2\, \partial_x
 \sum_{\alpha} e_{\alpha}  f^\alpha \right)
 \left( \int \textrm{d} v \sum_{\alpha}
 \frac{e_\alpha^2}{m_\alpha} f^\alpha  \right)^{-1} \,.
 \end{equation}
The initial conditions for system (\ref{bunch-kin}), (\ref{quasineutral}) correspond to the
distribution functions of electrons and ions at the instant $t = 0$:
 \begin{equation}\label{initial-cond}
 f^\alpha\big{|}_{ \, {t=0}} = f^\alpha_{0}(x,v)\,.
 \end{equation}
To construct the RG symmetry, we regard the system of local (\ref{bunch-kin}) and nonlocal
(\ref{quasineutral}) equations as the manifold $\cal{RM}$  (step \textbf{I} of the RG algorithm),
on which the electric field $E(t, x)$ is to be determined. The calculation of the Lie group of
point transformations admitted by this manifold (step \textbf{(II)} of the RG algorithm) defines a
finite-dimensional algebra generated by the operators of time and space shifts, the Galilean
transformation operator, three dilation operators, the quasineutrality operator, and the operator
of the projective group. The restriction of the group (step \textbf{(III)} of the RG algorithm) on
a particular solution of problem (\ref{bunch-kin}), (\ref{quasineutral}), (\ref{initial-cond}) with
a spatially symmetric initial distribution function with zero mean velocity selects a linear
combination of the time translation operator and the projective group operator. Under this
combination, the approximate solution of the initial problem $f^{\alpha}= f_0^{\alpha}(x,v) + O(t)$
provided by the perturbation theory as $t \to 0$ is invariant, and therefore this linear
combination is the RG symmetry operator:
 \begin{equation}  \label{rg-oper}
 R = (1 + \Omega^2 t^2 ) \partial_t  + \Omega^2 t x \partial_{x}
           + \Omega^2 (x-vt) \partial_{v} \, .
\end{equation}
The constant $\Omega\,$ can be regarded as the ratio of the characteristic sound speed $c_s$ to the
initial inhomogeneity scale of the plasma density $L_0$.\par

The invariants of RG operator (\ref{rg-oper}) are given by the distribution functions of the
particles $f^\alpha$ and the combinations $x/\sqrt{1+\Omega^2 t^2}$ and $ {{v}}^2 + \Omega^2({x}-
{v} t)^2 $. Hence, the construction of a solution of the BVP (step \textbf{(IV)} of the RG
algorithm) reduces to expressing the distribution functions at an arbitrary instant $t\neq 0$ in
terms of initial data (\ref{initial-cond}) with the help of these invariants,
 \begin{equation} \label{invariant}
 f^\alpha = f^\alpha_0(I^{(\alpha)})\,, \
 I^{(\alpha)} = \frac{1}{2}\left( {{v}}^2 + \Omega^2({x}- {v} t)^2 \right)
         + \frac{e_\alpha}{ m_\alpha} \Phi_0(x') \,,
 \end{equation}
where the dependence of $\Phi_0\,$ on $x' = x / \sqrt{1+\Omega^2 t^2}$ is determined by
quasineutrality conditions (\ref{quasineutral}). A concrete example illustrating these formulas for
a plasma layer formed by a group of hot and cold electrons and two kinds of ions can be found in
\cite{kov-jetf-02}.\par

Applications of the RG symmetry operator are not confined to the construction of solutions of an
initial value problem for Eqns (\ref{bunch-kin}), (\ref{quasineutral}) or to finding the
corresponding distribution functions of particles. For practical purposes, more rough
characteristics of the plasma dynamics is often needed, for instance, the density of the particles
(ions) of a certain kind $n^q(t,x)$, which can be found by integrating the distribution function:
\begin{equation}
\label{integral-def}
 \begin{aligned}  &
  n^q(t,x)  =  \int\limits_{-\infty}^{\infty}\,{\rm d} v f^q(t,x,v)  \,.
  \end{aligned}
   \end{equation}
Straightforward integration of the distribution function with respect to the velocity cannot always
be performed analytically because this function may have a complicated dependence on the invariant
$I^{(\alpha)}$. In this case, we can use the extension of the RG symmetry operator to a functional
of the solution because the density $n^q(t,x)$ is a linear functional of $f^q$. The extension of
operator (\ref{rg-oper}) to functional (\ref{integral-def}) yields the following operator in the
narrowed space of the variables $\{t,x,n^q\}$:
 \begin{equation} \label{rg-n}
 R = (1 + \Omega^2 t^2 ) \partial_t + \Omega^2 t x \partial_{x}
           - \Omega^2 t n^q \partial_{n^q}\,.
 \end{equation}
The solution of the Lie equations for operator (\ref{rg-n}) with initial conditions
(\ref{initial-cond}) taken into account yields a relation between the invariants of this operator
(one of the invariants, $J_{3} = x/\sqrt{1+\Omega^2 t^2}$, coincides with the above-mentioned
invariant of operator (\ref{rg-oper}), and the other invariant is $J_{4}^q = n^q \sqrt{1+\Omega^2
t^2}$ at an arbitrary instant $t \neq 0$ to their values at the initial instant $t=0$:
$J_{3}\!\!\mid _{t=0} = x^{\prime}$, $J_{4}^q\!\!\mid _{t=0} = {\cal N}_{q}( x^{\prime})$. This
relation immediately gives formulas describing the space-time distribution of the density of the
ions of given species in terms of their initial density distribution:
 \begin{equation} \label{density}
  \begin{aligned} &
  n^{q} = \frac{1}{\sqrt{1 + \Omega^2 t^2}}\, {\cal
  N}_{q}\left( \frac{x}{\sqrt{1+\Omega^2 t^2} } \right) \, , \quad
  {\cal N}_q(x^\prime ) = \int\limits_{-\infty}^{\infty}{\rm d} v f^q_0(I^{(q)})  \,.   \\
   \end{aligned}
\end{equation}
We note that the function ${\cal N}_q$ also characterizes the energy spectral distribution of ions
for large times $\Omega^2 t^2 > 1$ \cite{kov-jetf-02}. Thus, the use of the RG algorithm not only
allows constructing a solution of problem (\ref{bunch-kin}), (\ref{quasineutral}),
(\ref{initial-cond}) for various initial distribution functions of particles \cite{kov-jetf-02} but
also permits finding the law of the evolution of their density and their energy spectrum without
calculating the distribution functions of the particles explicitly. Similar results are obtained
not only in the framework of the model of a planar one-dimensional expansion but also, for
instance, for a spherically symmetric expansion of a plasma bunch \cite{kov-prl-03}.
\par

\section{Conclusion}\label{conclusion}

We now expound on several important points related to the development and applications of the RG
algorithm to BVPs of mathematical physics. \par

First of all, we note its universality, meaning that the procedure for the construction and the use
of RG symmetries is implemented in accordance with the scheme described in
Section~\ref{illustration}. Of course, approaches to the realization of the steps of the algorithm
can be different depending on the type of the problem under consideration, but the general pattern
of four successive steps remains the same. Our method not merely allows reproducing already known
solutions in a regular fashion but also produces new solutions.
\par

Second, the above examples do not exhaust all possible ways of implementing the RG algorithm. There
is an especially large freedom at the first step, that is, in the construction of the original
manifold. We have restricted ourself to the description of the most typical approaches (extending
the list of independent variables, using higher symmetries, applying the techniques of approximate
symmetries). We left the detailed description of the construction of the basic manifold with the
use of additional differential constraints and methods for the derivation of these constraints on
the basis of higher symmetries \cite{kov-jmp-98} outside the scope of this paper. One special case
of a differential relation defining a boundary condition is the embedding equation; this is
particularly interesting in mathematical models based on ordinary differential equations for which
the problem of symmetry calculation is nontrivial \cite{shkkp-rg-91,kov-jmp-98,ksh-phr-01}. We also
left the use of multiparameter renormalization groups \cite{shkkp-rg-91}, the construction of
approximate RG symmetries involving a small parameter in the transformation \cite{kov-tmf-99}, and
integration with respect to the RG-transformation parameter \cite{shkkp-rg-91} without detailed
speculation. For a detailed discussion of these issues and for applications of RG symmetries, the
reader can consult reviews ~\cite{kov-jmp-98,ksh-phr-01,ksh-proc-04,ksh-rg-05,shk-mathphys-05,
sh-lectures-92} and the references therein. \par

Third, we note that methods of computer algebra can be used for the construction of RG symmetries.
In the framework of the general scheme of the RG algorithm, one of the central computational
procedures is finding a maximal symmetry group of the manifold $\cal{RM}$. Here, it is necessary to
derive and solve a system of determining equations, which are linear (in the coordinates of the
infinitesimal operators) ordinary or partial differential equations. This usually amounts to
routine calculations, the bulk of which becomes quite large for higher-order symmetries and which
cannot be performed 'by hand' in a reasonably short time; psychologically, this can be a factor
constraining the use of the RG algorithm. However, by using methods of computer algebra at the
second step of the algorithm, often allows considerably accelerating the construction of RG
symmetries, as was shown in the example of the calculation of RG symmetries for equations of
quasi-Chaplygin media \cite{kpu-mcm-97}. The prospective gains can be at their greatest if analytic
and symbolic calculations are combined, when \textit{a priori} information about the form of the RG
symmetry extracted from analytic investigations can considerably reduce the time required for
symbolic calculations. Methods of symbolic calculations can be used for exact and approximate RG
symmetries alike, which significantly enhances the potentialities of the RG algorithm in general.
At the same time, analytic approaches used in constructing RG symmetries can be helpful in the
development of new algorithms for computer algebra systems. \par

Finally, we indicate possible ways to extend the scope of applications of the RG algorithm. This
can be achieved by covering new objects for which the use of the RG algorithm is not yet standard
or by modifying the algorithm itself. \par

One example of a new object can be an infinite system of coupled integrodifferential equations
similar to systems for correlation functions in statistical physics or to systems of equations for
generalized Green's functions, propagators, and vertex functions in quantum field theory. \par

As concerns modifications of the algorithm, they are connected in a natural way with the general
progress in the modern group analysis. This is how it became possible to extend the RG algorithm
(developed originally for physical problems described by differential equations) to nonlocal
problems. Certain hopes in this direction are related to the progress in group analysis in
application to generalized functions \cite{aks-dan-95}, further developing the theory of
approximate symmetries \cite{bai-nd-00}, finding new relations between the concept of symmetry and
conservation laws \cite{ibr-jmaa-06}, progress of the theory of partial-invariant solutions
\cite{ovs-dan-95,ovs-pmtf-95}, and applying group analysis to difference \cite{dor-bk-00} and
functional \cite{mel-cnsns-04} equations. \par

This research was carried out with the financial support of the ISTC (grant 2289), RFBR (grants
06-02-16103 and 08-01-00291), and LSS-1027.2008.2.

\end{document}